\documentclass[reprint,superscriptaddress,amsmath,amssymb,aps,prx]{revtex4-2}

\usepackage{newtxtext} 
\usepackage{newtxmath} 
\usepackage{comment}
\usepackage{enumitem}
\usepackage{graphicx}
\usepackage{dsfont}
\usepackage{dcolumn}
\usepackage{bm}
\usepackage{xcolor}
\usepackage{nameref}
\usepackage{tikz}
\usetikzlibrary{quantikz2}
\usepackage{hyperref}

\definecolor{grape}{RGB}{63, 37, 110}
\definecolor{ocean}{RGB}{29, 99, 174}
\hypersetup{
    colorlinks=true,
    linkcolor=ocean,
    filecolor=ocean,      
    urlcolor=ocean,
    citecolor=ocean
    }

\newcommand{\g}{\text{g}}
\newcommand{\q}{\text{q}}
\newcommand{\gr}{\text{grav}}
\newcommand{\br}{\text{bar}}
\def\id{\mathds{1}} 
\newcommand{\pureket}[1]{|{#1}\rangle\!\langle{#1}|}
\newcommand{\haf}{\text{haf}}
\newcommand{\lhaf}{\ell\text{haf}}
\newcommand{\tr}{\text{tr}}
\newcommand{\dd}{\text{d}}

\begin{document}

\title{Quantum State Characterization of Gravitational Waves via Graviton Counting Statistics}

\author{Kristian Toccacelo}
\email{kristo@dtu.dk}
\affiliation{Center for Macroscopic Quantum States (bigQ), Department of Physics, Technical University of Denmark, 
Fysikvej 307, 2800 Kongens Lyngby, Denmark}

\author{Thomas Beitel}
\affiliation{Department of Physics, 
Stevens Institute of Technology, 
Hoboken, New Jersey 07030, USA}

\author{Ulrik Lund Andersen}
\affiliation{Center for Macroscopic Quantum States (bigQ), Department of Physics, Technical University of Denmark, 
Fysikvej 307, 2800 Kongens Lyngby, Denmark}                            
\author{Igor Pikovski}
\email{pikovski@stevens.edu}
\affiliation{Department of Physics, 
Stevens Institute of Technology, 
Hoboken, New Jersey 07030, USA}
\affiliation{Department of Physics, 
Stockholm University, SE-106 91 Stockholm, Sweden}                    

\begin{abstract}
Although gravitational waves are now routinely observed, the detection of individual gravitons has long been regarded as impossible. Recent work, however, has demonstrated that single-graviton detection can be achieved and may be feasible in the near future. Here we show that beyond mere particle detection, these detectors provide access to the quantum state and particle statistics of gravitational waves. We show that graviton detection probabilities enable the discrimination between squeezed, coherent, and thermal radiation. We further demonstrate that the full quantum statistics contained in the second-order correlation function of the passing wave can be directly measured at the detector, independent of the weak gravitational interaction strength. Building on recent quantum-optical techniques, this capability opens the way to full quantum state tomography of Gaussian states. Our results demonstrate that single-graviton detection is not only of foundational significance but also of practical value, allowing for the characterization of quantum statistics and the states of the gravitational radiation field, which remain currently unknown.

\end{abstract}

\maketitle

\section{Introduction}
\label{sec:introduction}
Whether and how gravity is quantized remains one of the most fundamental open questions in physics. While general relativity successfully describes gravity at macroscopic scales, its reconciliation with quantum mechanics has been the subject of theoretical studies for nearly a century \cite{bronstein1936quantentheorie, rovelli2004quantum,oriti2009approaches}. For a long time, it was assumed that no realistic experiments could probe its predictions, most notably the existence of the graviton. Such a particle would presumably accompany any quantum theory of gravity and mediate the gravitational interaction. Most directly, the on-shell graviton is the quantum particle of gravitational waves (GWs), carrying discrete energy $E=hf$, where $h$ is the Planck constant, and $f$ the frequency of the wave. Until recently, it was assumed that direct detection of gravitons would be out of reach of experiments \cite{rothman2006can,dyson2013graviton}.  However, recent work by some of us has shown that gravitons can, in fact, be detected \cite{tobar2024} through absorption and detection of \textit{single} quanta of energy from passing gravitational waves. The result is possible due to now routine LIGO detections of passing GWs to which a single particle detector could cross-correlate, and recent advancements in the cooling and measurement of individual energy transitions of acoustic modes in macroscopic systems \cite{cohen2015phonon,hong2017,patil2022measuring,chu2017quantum,von2022parity,valimaa2022multiphonon,omahen2025ultra}. Combining these, it was demonstrated that graviton detection is realistic with near-future technology \cite{tobar2024}. 

\begin{figure}[t]
    \includegraphics[width=1\linewidth]{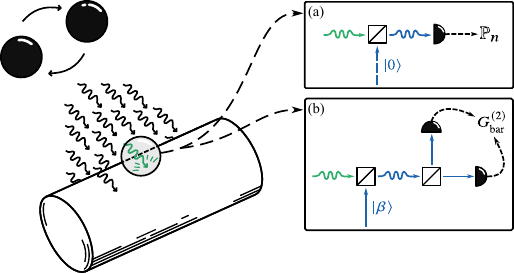}
    \caption{ 
    A gravitational wave (GW) from an astrophysical source is detected with a bulk acoustic resonator. The GW leaves an imprint in the form of resonant graviton-to-phonon conversion. (a) and (b) represent different detection schemes. (a) \textit{Measurement of single phononic excitations.} Different GW states result in different detection probabilities, opening the possibility for distinguishing different quantum states of the gravitational radiation. (b) Measurement of the second-order coherence. Detection of the phonon second-order correlation function in the bar detector gives direct information about the graviton statistics of the GW, which can be used to perform GW state tomography as the phononic initial state $\ket{\beta}$ is varied. }  
    \label{fig:graviton_scheme}
\end{figure}

Here, we show that, beyond mere detection, single-graviton detectors can also provide new input in our understanding of the quantum composition of gravitational waves. Quantum theory predicts that gravitational waves, like their electromagnetic counterparts, could come in a variety of quantum states, ranging from classical-like coherent states to highly non-classical states, like squeezed states, or even non-Gaussian states with negative Wigner functions and sub-Poissonian counting statistics. The non-linear nature of gravity and the GW production process indicates that passing GWs may indeed deviate from coherent states \cite{guerreiro2025entanglement, kanno2025quantum, manikandan2025squeezed, das2025squeezed}. However, identifying which states can be distinguished observationally remains an open challenge. The classical position detection method used for GW detection has limited sensitivity to features that determine the quantum state---unless it is exponentially enhanced \cite{parikh2021signatures,parikh2021quantum,kanno2021noise,kanno2022conversion,abrahao2023quantum}---and mostly confirms the classical mean corresponding to the intensity. But the same intensity can be produced by a variety of quantum states.  In particular, Gaussian states, characterized by a positive Wigner function and fully described by their first and second moments, are a wide class of different possible states, which play a central role in quantum optics and continuous-variable quantum information \cite{serafini2023quantum, braunstein2005quantum, brask2021gaussian}. It remains an open question what exact states of gravitational radiation astrophysical sources may produce, and which of these states may be detectable on Earth.

In this work, we propose a robust method to determine gravitational wave states through graviton counting statistics in graviton-click detectors. By analyzing the statistical properties of phonon number distributions in the detectors, we derive experimental criteria that distinguish Gaussian characteristics of the passing GWs and can even provide their full quantum statistics and state tomography. This framework provides a potential diagnostic for detecting the statistical particle nature of gravitational radiation and establishes a close link between graviton detection and modern methods in quantum optics.

\section{Interaction Hamiltonian and equations of motion}
\label{sec:interaction_hamiltonian_and_equations_of_motion}
We first briefly summarize the interaction dynamics between GWs and a resonant detector that leads to the ability to detect gravitons---see Ref.~\cite{tobar2024} and App.~\ref{app:interaction_hamiltonian_derivation}. We consider a gravitational wave incident on a macroscopic bulk acoustic resonator, which we assume to be cylindrical, of length $L$, and total mass $M$. The interaction Hamiltonian between the quantized modes of the gravitational radiation field $\hat{a}$ and $\hat{a}^\dagger$, and the collective phononic modes $\hat{b}_\ell$ and $\hat{b}_\ell^\dagger$, where $\ell$ is an odd integer, and $\omega_\ell$ the corresponding resonator frequency, can be written as \cite{grishchuk1992quantum,pang2018quantum,tobar2024,manikandan2025complementary,loughlin2025wave}
\begin{align}
\label{eq:interaction_hamiltonian}
     \hat{H}_\text{int}=\hbar\, \gamma_\g\left(\hat{a}\, e^{-i \nu t}+\hat{a}^\dagger e^{i \nu t} \right)\left(\hat{b}_\ell+\hat{b}^\dagger_\ell\right).
\end{align}
Here, $\nu=\nu(t)$ is the gravitational wave frequency, $\gamma_\g$ is the coupling strength given by
\begin{align}
\label{eq:coupling_strength}
    \gamma_\g=\sqrt{(-1)^{\ell-1}\frac{8\pi G M\nu^3L^3}{\omega_\ell\, c^2 V \pi^4 \ell^4}},
\end{align} 
$G$ is Newton's constant, $c$ the speed of light in vacuum, $V$ the characteristic volume of the gravitational wave, and $\omega_\ell$ the frequency of the resonator. In writing \eqref{eq:interaction_hamiltonian}, we assume a large GW wavelength compared to the detector and that the interaction involves a single mode of the gravitational field, that is, the mode on resonance with the acoustic resonator at time $t$ (in general, the GW frequency will chirp but is still well described by a single mode). Given that the coupling is typically smaller than the resonator's frequency, it is possible to further approximate the interaction Hamiltonian by employing the rotating wave approximation (RWA), in which case
\begin{align}
\label{eq:interaction_hamiltonian_rwa}
     \hat{H}_\text{int}=\hbar\, \gamma_\g\left(\hat{b}^\dagger_\ell\,\hat{a}\, e^{-i \nu t}+\hat{b}_\ell\,\hat{a}^\dagger e^{i \nu t} \right).
\end{align}
The dynamics generated by the Hamiltonian \eqref{eq:interaction_hamiltonian_rwa} are easily solved in the Heisenberg picture, where we can view the time dependence of the mode operators as the beamsplitter transformations
\begin{align}
\begin{gathered}
\label{eq:gravitational_beam_splitter}
    \hat{b}_\ell(t) = e^{-i \omega_\ell t} \left[ \cos(\gamma_\g t)\,\hat{b}_\ell-i \sin(\gamma_\g t)\,\hat{a}  \right], \\
    \hat{a}(t) =  e^{-i \omega_\ell t}\left[\cos(\gamma_\g t)\,\hat{a} -i \sin(\gamma_\g t)\,\hat{b}_\ell  \right], 
\end{gathered}  
\end{align}
where $\hat{b}_\ell=\hat{b}_\ell(0)$ and $\hat{a}=\hat{a}(0)$. The solutions to the equations of motion given by \eqref{eq:gravitational_beam_splitter} hold when the resonator to GW detuning parameter, defined as $\Delta=\omega_\ell-\nu$, vanishes; in the Apps.~\ref{app:evolution_in_the_rotating_wave_approximation}~and~\ref{app:evolution_beyond_the_rotating_wave_approximation}, we respectively show the solutions for $\Delta\neq0$, and the solutions with the RWA condition relaxed. Given the initial state of the detector and the gravitational wave modes, the solutions in Eq.~\eqref{eq:gravitational_beam_splitter} allow us to determine the expectation values of any observable of interest, for any Gaussian or non-Gaussian initial state.

\subsection{Dynamics for general Gaussian gravitational wave states}
In Ref.~\cite{tobar2024}, a coherent GW state was assumed, and it was shown that measurement of $\hat{b}^\dagger_\ell \hat{b}_\ell$ yields direct information about single graviton absorption and emission. In the following, we will be interested in characterizing the dynamics for general Gaussian states and other observables. Since we are in a regime where the interaction between the two modes is quadratic in the mode operators [see Eq.~\eqref{eq:interaction_hamiltonian}], all states that are initially Gaussian will evolve into final Gaussian states. This remains true when tracing out any subsystem, for instance, the reduced state $\rho^{}_\br(t)$ of the detector system, which will be relevant for us in the following discussion. In general, an $N$-mode Gaussian state with quadrature operators $\hat{\boldsymbol{r}}=(\hat{x}_1,\hat{p}_1,\dots,\hat{x}_N,\hat{p}_N)^\text{T}$, is completely characterized by the first and second moments, namely, by the covariance matrix $\boldsymbol{\sigma}$ and displacement vector $\bar{\boldsymbol{r}}$ defined by
\begin{align}
\label{eq:canonical_covariance_matrix_displacement_vector}
    \sigma_{ij}=\frac{1}{2}\langle\{\hat{r}_i,\hat{r}_j\}\rangle-\langle \hat{r}_i\rangle\langle\hat{r}_j\rangle,\,\text{and}\quad \bar{r}_i=\langle \hat{r}_i\rangle,
\end{align}
where, given the operator $\hat{A}$, the expectation value is $\langle \hat{A}\rangle=\tr[\rho \hat{A}]$. The evolution of Gaussian moments under a quadratic Hamiltonian is determined by the symplectic transformations
\begin{align}
\label{eq:symplectic_evolution}
    \bm{\sigma}(t)=S\boldsymbol{\sigma}(0) S^\text{T}, \quad \bar{\bm{r}}(t)=S\,\bar{\boldsymbol{r}}(0),
\end{align}
where $S=e^{\boldsymbol{\Omega}  Ht/\hbar}$, $\boldsymbol{\Omega}$ is the symplectic form, and $H$ is the Hamiltonian matrix 
\begin{align}
    \hat{H}=\frac{1}{2}\hat{\boldsymbol{r}}^\text{T}H\hat{\boldsymbol{r}}+\hat{\boldsymbol{r}}^\text{T}\bar{\boldsymbol{r}}.
\end{align}
Using this formalism, let us consider the dynamics of a general Gaussian gravitational wave state $\rho_\gr$ interacting with a resonant detector prepared in the ground state $\rho_\br=\pureket{0}$. In this case, the gravitational wave state can be written as a combination of the squeezing and displacement operators acting on a thermal state for a certain choice of displacement $\alpha\in\mathbb{C}$, squeezing $\xi\in\mathbb{C}$, and average thermal occupation number $\bar{n}\in\mathbb{R}$. It follows that an arbitrary Gaussian gravitational wave state can be cast in the form
\begin{align}
\label{eq:general_gaussian_gravitational_wave_state}
    \rho^{}_\gr =  \hat{D}_\alpha\,\hat{S}_\xi\,\rho^{}_{\bar{n}}\,\hat{S}^\dagger_\xi\,\hat{D}^\dagger_\alpha,
\end{align}
where
\begin{align}
    \hat{S}_\xi=e^{\frac{1}{2}\left({\xi^*}\hat{a}^2-{\xi}\,{\hat{a}^{\dagger2}}\right)}, \quad \hat{D}_\alpha=e^{\alpha\, \hat{a}^\dagger-\alpha^* \hat{a}},
\end{align}
are the squeezing and displacement operators, respectively, and $\rho^{}_{\bar{n}}$ is the thermal state
\begin{align}
    \rho^{}_{\bar{n}}=\frac{1}{1+\bar{n}}\sum_{n=0}^{\infty}\left(\frac{\bar{n}}{\bar{n}+1}\right)^n|n\rangle\!\langle n|.
\end{align} 
At the level of the covariance matrix and displacement vector, the general state \eqref{eq:general_gaussian_gravitational_wave_state} can be written as
\begin{align}
    \boldsymbol{\sigma}_\gr(0)=\boldsymbol{F}_\xi\,\boldsymbol{\sigma}_{\bar{n}}\,\boldsymbol{F}_\xi^\text{T},\quad \bar{\boldsymbol{r}}_\gr(0)=\sqrt{2}\left(\text{Re}(\alpha),\text{Im}(\alpha)\right)^\text{T},
\end{align}
where $\boldsymbol{F}_\xi=\cosh r \,\id_2-\sinh r \,\mathfrak{R}_\varphi$ is the squeezing transformation, $\xi=r e^{i \theta}$, $\mathfrak{R}_\theta$ is the matrix
\begin{align}
    \mathfrak{R}_\varphi=\begin{pmatrix}
        \cos\theta & \sin\theta \\
        \sin\theta & -\cos\theta
    \end{pmatrix},
\end{align} 
and $\boldsymbol{\sigma}_{\bar{n}}$ the covariance matrix of a thermal state with average thermal occupation $\bar{n}$, given by
\begin{align}
    \boldsymbol{\sigma}_{\bar{n}}=\begin{pmatrix} \frac{1}{2}+\bar{n} & 0 \\ 0 & \frac{1}{2}+\bar{n} 
    \end{pmatrix}.
\end{align}
For the detector initialized in the vacuum state, the state at $t=0$ is specified by
\begin{align}
    \boldsymbol{\sigma}_\br(0)=\frac{1}{2}\id_2,\quad \bar{\boldsymbol{r}}_\br(0)=0.
\end{align} 
The joint state of the gravitational wave and detector system after having resonantly interacted through the Hamiltonian in Eq.~\eqref{eq:interaction_hamiltonian_rwa}, can be determined via the corresponding symplectic transformations \eqref{eq:symplectic_evolution}---see App. \ref{app:gaussian_evolution} for details. In particular, the reduced state of the detector at time $t$ is given, up to a local phase shift, by the first and second moments
\begin{align}
\label{eq:gaussian_state_x_p}
\begin{gathered}
    \boldsymbol{\sigma}_\br(t)= \cos^2(\gamma_\g t) \boldsymbol{\sigma}_\br(0)+\sin^2(\gamma_\g t)\mathcal{}\boldsymbol{\sigma}_\gr(0),\\ \bar{\boldsymbol{r}}_\br(t)=\sin(\gamma_\g t)\,\bar{\boldsymbol{r}}_\gr(0).
\end{gathered}
\end{align}
For what we are about to consider, it is convenient to represent the first and second moments of the Gaussian state using the creation and annihilation operators. For $N$ modes, we define the ladder operator
\begin{align}
\label{eq:cration_and_annihilation}
    \hat{a}_j=\frac{1}{\sqrt{2}}\left(\hat{x}_j+i\hat{p}_j\right),\quad\hat{a}^\dagger_j=\frac{1}{\sqrt{2}}\left(\hat{x}_j-i\hat{p}_j\right),
\end{align}
which, like the canonical quadrature operators, can be collected in the column vector $\hat{\boldsymbol{a}}=(\hat{a}_1,\hat{a}_1^\dagger,\dots,\hat{a}_N,\hat{a}_N^\dagger)^\text{T}$. In this representation, a Gaussian state is characterized by the first and second moments
\begin{align}
\label{eq:rotated_covariance_matrix_displacement_vector}
    \Sigma_{ij}=\frac{1}{2}\langle\{\hat{{a}}_i,\hat{{a}}_j\}\rangle-\bar{a}_i\bar{a}_j^*,\quad\bar{a}_i=\langle\hat{a}_i\rangle.
\end{align}
We can go from the representation in Eq.~\eqref{eq:canonical_covariance_matrix_displacement_vector} to the representation in Eq. \eqref{eq:rotated_covariance_matrix_displacement_vector} through the linear transformation
\begin{align}
    \bar{\boldsymbol{a}}=\boldsymbol{M}\,\bar{\boldsymbol{r}}, \quad \boldsymbol{\Sigma}=\boldsymbol{M}\,\boldsymbol{\sigma}\,\boldsymbol{M}^\dagger
\end{align}
where $\boldsymbol{M}$ is defined by
\begin{align}  
\label{eq:quadrature_rotation}\boldsymbol{M}=\bigoplus_{j=1}^{N} M_j, \quad\text{with}\quad M_j=\frac{1}{\sqrt{2}}\begin{pmatrix}
        1 &i\\1 &-i
    \end{pmatrix} .
\end{align}

\section{Graviton counting statistics}
\label{sec:graviton_counting_statistics}
The beamsplitter transformations \eqref{eq:gravitational_beam_splitter} tell us that if the incident gravitational wave is in a coherent state $|\alpha\rangle_\gr$ and the detector is prepared in its ground state $|0\rangle_\br$, the joint state at time $t$ is given by
\begin{align}
    \hat{U}_t\left(|\alpha\rangle_\gr\otimes|0\rangle_\br\right)=|\alpha \cos(\gamma_\g t)\rangle \otimes |\!-i\alpha \sin(\gamma_\g t)\rangle,
\end{align}
where $\hat{U}_t=\exp\left(-\frac{i\,t}{\hbar}\hat{H}\right)$ is the time-evolution operator. In this case, since the interaction with the gravitational wave induces a coherent state in the detector system, the probability $\mathbb{P}_n$ of detecting $n$ GW-induced excitations in the detector is simply the Poisson distribution
\begin{align}
\label{eq:coherent_state_p_n}
    \mathbb{P}_n(t)=e^{-|\alpha|^2\sin^2(\gamma_\g t)}\frac{\left(|\alpha|^2\sin^2(\gamma_\g t)\right)^n}{n!},
\end{align}
which comes about when projecting the state of the detector onto the $n$-th Fock state $|n\rangle$. This is the basis of single-graviton detection \cite{tobar2024}, as the detection of one excitation in the detector corresponds to the resonant swap of exactly one quantum from the GW. When the gravitational field is initially in the general Gaussian state $\rho_\gr$ given by Eq.~\eqref{eq:general_gaussian_gravitational_wave_state}, the reduced state of the detector after interacting with the GW is not necessarily pure. Therefore, determining $\mathbb{P}_n$ by projecting onto the Fock state $|n\rangle$ like above is not trivial. Instead, we resort to the phase space representation. In general, for an arbitrary state $\rho$ the probability $\mathbb{P}_n=\tr[\rho \, \hat{m}_n]$, where we defined the fock state projector $\hat{m}_n=\pureket{n}$, can be written as the overlap integral
\begin{align}
\label{eq:phase_space_probability}
    \mathbb{P}_n=\int \dd^2 \boldsymbol{\alpha}\, Q_\rho(\boldsymbol{\alpha})P_n(\boldsymbol{\alpha}).
\end{align}
Here, $Q_\rho(\boldsymbol{\alpha})=\langle \alpha|\rho|\alpha\rangle/\pi$ is the Husimi $Q$-function of the state $\rho$, $P_n(\boldsymbol{\alpha})$ is the Glauber–Sudarshan $P$-representation of the projector $\pureket{n}$, and $\boldsymbol{\alpha}=(\alpha,\alpha^*)$.

\subsection{Counting excitations in Gaussian states} 
 To explicitly determine the probabilities \eqref{eq:phase_space_probability}, we use the representation of the first and second moments in terms of $\mathbf{\Sigma}$ and $\bar{\mathbf{a}}$ defined in \eqref{eq:rotated_covariance_matrix_displacement_vector}. The $Q$-function is then represented by
\begin{align}
\label{eq:q_function_definition}    
    Q^{}_\rho(\boldsymbol{\alpha})=\frac{1}{\sqrt{\det(\boldsymbol{\Sigma_Q})}}e^{-\left(\boldsymbol{\alpha}-\bar{\mathbf{a}}\,\right)^\dagger\boldsymbol{\Sigma_Q}^{-1}\left(\boldsymbol{\alpha}-\bar{\mathbf{a}}\,\right)/2},
\end{align}
where $\boldsymbol{\Sigma_Q}=\boldsymbol{\Sigma}+\frac{1}{2}\id$. The $P$-representation of the Fock state $\pureket{n}$, on the other hand, is singular, but can be represented using Dirac delta functions, namely,
\begin{align}
\label{eq:n_fock_state_p_function}P_n(\boldsymbol{\alpha})=\frac{e^{|\alpha|^2}}{n!}\left(\frac{\partial^2}{\partial\alpha\,\partial\alpha^*}\right)^n\delta(\alpha)\,\delta(\alpha^*).
\end{align}
If we plug Eqs.~\eqref{eq:q_function_definition} and \eqref{eq:n_fock_state_p_function} into Eq.~\eqref{eq:phase_space_probability}, we can show that the probabilities are expressed in terms of a generating function \cite{kruse2019a}, namely,
\begin{align}
\mathbb{P}_n=\left.\left(\frac{\partial^2}{\partial \alpha \partial \alpha^*}\right)^n \mathcal{G}(\alpha,\alpha^*)\right|_{\boldsymbol{\alpha}=0},
\label{eq:probabilities_from_generating_function}
\end{align}
where the generating function is given by 
\begin{align}
\label{eq:probability_generating_function}
    \mathcal{G}(\alpha,\alpha^*)=\frac{e^{-\frac{1}{2}
\bar{\boldsymbol{a}}^{\dagger} \boldsymbol{\Sigma_Q}^{-1} \bar{\boldsymbol{a}}}}{n!\sqrt{\det(\boldsymbol{\Sigma_Q})}}
\exp\left[{\frac{1}{2}\boldsymbol{\alpha}^\text{T} \boldsymbol{A} \boldsymbol{\alpha}+\boldsymbol{F}} \boldsymbol{\alpha}\right],
\end{align}
and where we defined 
\begin{align}
    \boldsymbol{A}=\begin{pmatrix}
        0 & 1\\
        1 & 0
    \end{pmatrix} \left(\id-\mathbf{\Sigma}_Q^{-1}\right),\quad \boldsymbol{F}=\bar{\boldsymbol{a}}^\dagger\boldsymbol{\Sigma}_Q^{-1}.
\end{align}
We note that the probabilities \eqref{eq:probabilities_from_generating_function} can also be expressed in terms of the so-called loop Hafnian~\cite{kruse2019a}, namely,
\begin{align}
\label{eq:loop_hafnian_detection_probabilities}
    \mathbb{P}_n=\frac{e^{-\frac{1}{2}
\bar{\boldsymbol{a}}^\dagger \boldsymbol{\Sigma}_Q^{-1} \bar{\boldsymbol{a}}}}{n!\sqrt{\det(\boldsymbol{\Sigma}_Q)}}\lhaf(\boldsymbol{\mathcal{A}}^{(n)}),
\end{align}
where $\boldsymbol{\mathcal{A}}^{(n)}$ is matrix constructed from $\boldsymbol{A}^{(n)}=\boldsymbol{A}\otimes\id_n$ by replacing the diagonal entries with the $2n$-dimensional column vector $\boldsymbol{F}^{(n)}$ obtained repeating $n$ times the components of $\boldsymbol{F}=\bar{\boldsymbol{a}}^\dagger\,\boldsymbol{\Sigma}_Q^{-1}$, and $\lhaf$ is the loop hafnian, defined as
\begin{align}
    \ell\haf(B)=\sum_{p\in \text{P}^{1,2}_{2k}} \prod_{\{i,j\}\in p}B_{ij}\,,
\end{align}
$B$ being a $2k\times2k$ symetric matrix, and $\text{P}^{1,2}_{2k}$ being the set of partitions of size 1 and 2 of $2k$ elements \cite{bjorklund2019}.

\subsection{Graviton statistics of Gaussian gravitational wave states}
We now use the formalism outlined in the previous section to determine the transition probabilities $0\to n$ induced by the gravitational wave on the detector system. In the creation and annihilation operators representation, the reduced state of the bar resonator at time $t$ after having resonantly interacted with the incident Gaussian gravitational radiation is given by 
\begin{align}
\label{eq:gaussian_state_ladder_operator}
\begin{gathered}
    \boldsymbol{\Sigma}_\br(t)= \cos^2(\gamma_\g t) \boldsymbol{\Sigma}_\br(0)+\sin^2(\gamma_\g t)\boldsymbol{\Sigma}_\gr(0),\\ \bar{\boldsymbol{a}}_\br(t)=\sin(\gamma_\g t)\,\bar{\boldsymbol{a}}_\gr(0).    
\end{gathered}
\end{align}
From Eq.~\eqref{eq:loop_hafnian_detection_probabilities}, we find that the probability for the resonator to be found in the ground state at time $t$ is given by
\begin{align}
    \mathbb{P}_0(t)=\frac{\exp\left[-\frac{1}{2}
\bar{\boldsymbol{a}}_\br^\dagger (\boldsymbol{\Sigma}_\br-\frac{1}{2}\id_2)^{-1} \bar{\boldsymbol{a}}_\br\right]}{\sqrt{\det(\boldsymbol{\Sigma}_\br-\frac{1}{2}\id_2)}},
\end{align}
while for the transition $0\rightarrow1$ we find 
\begin{align}   \mathbb{P}_1(t)=\mathbb{P}_0(t)\left[\mathfrak{a}_1+\mathfrak{f}_1\mathfrak{f}_2\right]
\end{align}
and, similarly, for the transition $0\to2$ we have
\begin{align}
    \mathbb{P}_2(t)=&\frac{1}{2}\mathbb{P}_0(t)\left[\left(\mathfrak{a}_2+\mathfrak{f}_2^2\right)\left(\mathfrak{a}_2+\mathfrak{f}_1^2\right)+4 \mathfrak{f}_1 \mathfrak{f}_2 \mathfrak{a}_1+2\,\mathfrak{a}_1^2\right],
\end{align}
where $\mathfrak{a}_1$, $\mathfrak{a}_2$, $\mathfrak{f}_1$, and $\mathfrak{f}_2$, are defined in Eqs.~\eqref{eq:def_a_1}--\eqref{eq:def_f_2}, and depend on time, the displacement magnitude $|\alpha|$, the squeezing parameter $r$, the thermal occupation number $\bar{n}$, and the relative phase between the displacement and squeezing phases. Without loss of generality, we choose the relative phase so that it coincides with the squeezing angle $\theta$. Explicitly, in the case where $\theta=0$, we have
\begin{widetext}
\begin{align}
\begin{gathered}
    \mathbb{P}_0(t)=\frac{2 \exp \left\{-\frac{4 |\alpha|^2 e^{2 r} \sin ^2(\gamma_\g t)}{2 (2 \bar{n}+1) \sin ^2(\gamma_\g t)+e^{2 r} [\cos (2 \gamma_\g t)+3]}\right\}}{\sqrt{(2 \bar{n}+1) \cosh (2 r) \sin ^2(\gamma_\g t) [\cos (2 \gamma_\g t)+3]+(2 \bar{n}+1)^2 \sin ^4(\gamma_\g t)+\left[\cos ^2(\gamma_\g t)+1\right]^2}},\\
    \mathbb{P}_1(t)= \mathbb{P}_0(t)\!\left\{\!1\!-\!\frac{2}{2 (2 \bar{n}+1) e^{2 r} \sin ^2(\gamma_\g t)+\cos (2 \gamma_\g t)+3}\!-\!\frac{4 (2 \bar{n}+1) e^{2 r} \sin ^2(\gamma_\g t)+2 e^{4 r} \left[\left(4 |\alpha |^2+1\right) \cos (2 \gamma_\g t)+3-4 |\alpha|^2\right]}{\left[2 (2 \bar{n}+1) \sin ^2( \gamma_\g t)+e^{2 r} (\cos (2 \gamma_\g t)+3)\right]^2}\!\right\}.    
\end{gathered}
\end{align}
\end{widetext}

\begin{figure*}[t]
\centering
\includegraphics[width=1\linewidth]{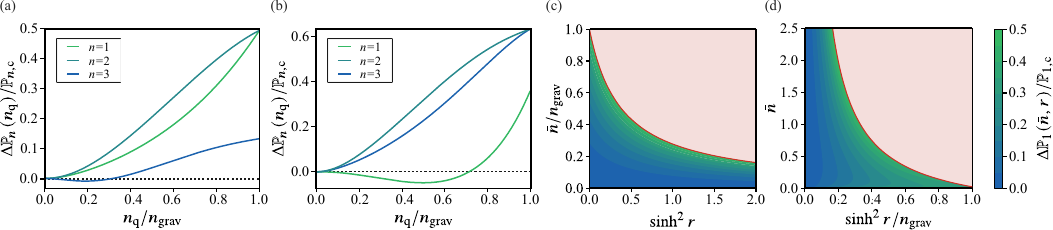}
\caption{Graviton absorption depends on the state of the passing gravitational wave. Here, we assume a passing GW with strain amplitude $h=10^{-22}$ and show the fractional graviton absorption probability difference between a purely coherent and general Gaussian gravitational wave state. $n_\q/n_\gr$ is the fraction of squeezed and thermal gravitons to the total intensity, as given by Eq.~\eqref{eq:average_graviton_count}. The curves in (a) and (b) show the first three excitation probabilities for a detector coupling strength of $n_\gr (\gamma_\g t)^2 \approx1$ and $n_\gr (\gamma_\g t)^2 \approx2$, respectively. (c) and (d) are contour plots of the fractional difference in single-graviton absorption $\Delta\mathbb{P}_1(\bar{n},r)/\mathbb{P}_{1,\text{c}}$ for the mostly thermal $n_\q \approx \bar{n}$ limit (c), and the mostly squeezed $n_\q \approx \sinh^2(r)$ limit (d), respectively, with the interaction strength as in (a). 
}
\label{fig:delta_p_figure}
\end{figure*}

Let us compare the excitation probability to the case of an exact coherent state input, as considered in the original work on single graviton detection \cite{tobar2024}. We are interested in determining the possible state of the GW while fixing the incoming graviton flux, given by the classically measured intensity through the GW strain amplitude $h$ with the relationship \cite{boughn2006aspects,dyson2013graviton}
\begin{equation}\label{eq:intensity}
    n_\text{grav} = \frac{h^2}{ 32 \pi \nu^2 t_\text{P}^2 },
\end{equation}
where $\nu$ is the GW frequency, and $t_\text{P} = \sqrt{\hbar G/c^5}$ the Planck-time. For a typical wave detected by LIGO with $h = 10^{-22}$ and $\nu = 2\pi \times 100\,\text{Hz}$, this corresponds to an average number of gravitons of $n_\text{grav} \approx 10^{35}$. For a general Gaussian state, it is given by 
\begin{equation} 
\label{eq:average_graviton_count}
n_\gr = |\alpha|^2+\left(\bar{n}+\frac{1}{2}\right) \cosh (2 r)-\frac{1}{2} \equiv  |\alpha|^2 + n_\q  \, ,
\end{equation}
where we defined $n_\q(\bar{n},r)$ as the average number of squeezed and thermal gravitons. As the detection probabilities depend on the GW state, we can consider the difference between the probabilities for a purely coherent GW input state ($n_\q=0$) and a general Gaussian state. In particular, we consider the quantity
\begin{align}
\label{eq:coherence_deviation}
    \Delta \mathbb{P}_n=\mathbb{P}_{n,{\text{c}}}-\mathbb{P}_n,
\end{align}
where $\mathbb{P}_{n,{\text{c}}}$ is simply the coherent state number distribution \eqref{eq:coherent_state_p_n}. To lowest order in $(\gamma_\g t)^2$, with $n_\text{grav} (\gamma_\g t)^2 $ kept finite, the single graviton deviation $\Delta\mathbb{P}_1$ is given by
\begin{align}
\label{eq:P-diff}
    \frac{\Delta \mathbb{P}_1(\bar{n}, r)}{\mathbb{P}_{1,{\text{c}}}} \! \approx \! 1 - \frac{2\left( 1 - \frac{n_\q}{n_\gr} \right) n_\q (\gamma_\g t)^2 + 1 }{\left[2 n_\q (\gamma_\g t)^2 +1 \right]^{3/2}} \, e^{n_\q(\gamma_\g t)^2}.
\end{align}
We plot the normalized single-graviton difference $\Delta\mathbb{P}_1/\mathbb{P}_{1,{\text{c}}}$ and its generalization to $\Delta\mathbb{P}_n/\mathbb{P}_{n,{\text{c}}}$ for different states of the GW in Figure \ref{fig:delta_p_figure}, showing what ranges of $n_\q$ could be detected. A state other than a purely coherent state thus imprints a signature that can, in principle, be detected when the absorption statistics are accurately measured. It is unlikely that the GW intensity is dominated by contributions other than from the coherent state. In the case $\vert\alpha\vert^2 \gg n_\q$, the expression \eqref{eq:P-diff} becomes $\Delta \mathbb{P}_1 \approx   \frac{1}{2} n_\gr \left[ (2 \bar{n}+1)e^{2r} - 1\right] (\gamma_\g t)^4 e^{-n_\gr (\gamma_\g t)^2} \approx   \frac{1}{2} n_\gr \, n_\q (\gamma_\g t)^4 e^{-n_\gr (\gamma_\g t)^2}$, where the last approximation is valid for $r \gtrsim 1$. Thus, even though we are in the limit $n_\gr (\gamma_\g t)^2 \approx 1 $, as required for the ability to detect graviton clicks, the change in the probabilities of detection due to GW states other than coherent states scales as $\Delta \mathbb{P}_1 \sim n_\q/ n_\gr$, and is, therefore, resolvable when $n_\q/ n_\gr$ is not too small. In fact, the authors of Ref.~\cite{manikandan2025squeezed} estimate that the number of squeezed quanta from compact binary mergers as detected on Earth could be relatively large, on the order of $n_\q/ n_\gr \sim 0.01$. 

While the above single-phonon excitation is well suited for $n_\gr (\gamma_\g t)^2 \approx 1 $, i.e., a detector that mostly measures single graviton clicks, one can similarly derive the detection probabilities for better detectors that absorb more than one graviton, using Eq.~\eqref{eq:loop_hafnian_detection_probabilities}---see also Fig.~\ref{fig:delta_p_figure} (b). In the following, we present a more general way to determine the quantum state of the GW based on counting statistics.

\section{Second-order correlations}
\label{sec:second_order_correlations}
In quantum optics, the statistics of a light field can be characterized by second-order correlation functions that measure correlations between intensities \cite{loudon2000quantum}. These tools are also applicable to the statistics of gravitational radiation. In this section, we will be interested in the zero-time-delay $g^{(2)}(0)$ function, defined as
\begin{align}
\label{eq:second_order_autocorrelation_function}
g^{(2)}(0)=\frac{\langle\hat{n}^2-\hat{n}\rangle}{\langle \hat{n}\rangle^2}.
\end{align}

We can use the same formalism detailed in the previous section for determining excitation probabilities of Gaussian states to determine the expectation values needed to compute the second-order coherence \eqref{eq:second_order_autocorrelation_function}. In particular, the $s$-ordered moments of an arbitrary state $\rho$ can be expressed in terms of derivatives of the $s$-ordered characteristic function $\chi_s(\boldsymbol{\alpha})=e^{s|\alpha|^2/4}\tr[\rho \hat{D}_{\boldsymbol{\alpha}}]$, namely \cite{barnett2002, cardin2024}
\begin{align}
\label{eq:s_ordered_moments}
    \langle\hat{a}^{\dagger n}\,\hat{a}^m\rangle_s=\left.\left(\frac{\partial}{\partial\alpha}\right)^n\left(-\frac{\partial}{\partial\alpha^*}\right)^m \chi^{}_s(\boldsymbol{\alpha})\right|_{\boldsymbol{\alpha}=0}.
\end{align}
Here, $s=1$ corresponds to normal ordering, $s=0$ corresponds to symmetric ordering, and $s=-1$ corresponds to ant-normal ordering. For a Gaussian state with covariance matrix $\boldsymbol{\Sigma}$ and displacement $\bar{\boldsymbol{a}}$, the characteristic function is given by 
\begin{align}
    \chi^{}_s(\boldsymbol{\alpha})=\exp\left({-\frac{1}{2}\boldsymbol{\alpha}^\dagger \boldsymbol{Z} \,\boldsymbol{\Sigma}_s \boldsymbol{Z}\,\boldsymbol{\alpha}+\bar{\boldsymbol{a}}^\dagger \boldsymbol{Z}\,\boldsymbol{\alpha}}\right),
    \end{align}
where we defined
\begin{align}
    \boldsymbol{\Sigma}_s=\boldsymbol{\Sigma}-\frac{s}{2}\id, \quad\text{and} \quad \boldsymbol{Z}=\begin{pmatrix}
        0 & 1 \\
        1 & 0
    \end{pmatrix}.
\end{align}
Notice that $\boldsymbol{\Sigma}_Q=\boldsymbol{\Sigma}_{-1}$. Indeed, the $Q$-function \eqref{eq:q_function_definition} is related to the anti-normal ordered characteristic function by a Fourier transform. As for the excitation probabilities discussed above, it is possible to relate the $s$-ordered moments to the loop-Hafnian of an appropriately defined matrix---see Ref.~\cite{cardin2024}.

\subsection{Measuring graviton counting statistics}
We consider the case in which the initial joint state of the gravitational wave and detector system is $\rho=\rho^{}_\gr\otimes|0\rangle\!\langle0|^{}_\br$. In this case, we have
\begin{align} \label{eq:excitations}
\langle\hat{n}^{}_\br(t)\rangle=\sin^2(\gamma_\g t) \langle \hat{n}^{}_\gr\rangle ,
\end{align}
and similarly
\begin{align}
\langle\hat{n}_\br^2(t)\rangle=\sin^4(\gamma_\text{g}t)\langle\hat{n}^2_\gr\rangle+\sin^2(\gamma_\text{g}t)\cos^2(\gamma_\text{g}t)\langle\hat{n}^{}_\gr\rangle.
\end{align}
Consequently, for the $g^{(2)}(0)$ function we find
\begin{align}
\label{eq:general_bar_second_order_zero_time_delay}
    g_\br^{(2)}(0)&=\frac{\langle\hat{n}^2_\gr\rangle-\langle \hat{n}^{}_\gr\rangle }{\langle \hat{n}^{}_\gr\rangle^2}\equiv g_\gr^{(2)}(0),
\end{align}
which is valid for arbitrary states of the gravitational field $\rho_\gr$, and for $\sin(\gamma_g t)\ne 0$. 
Remarkably, we see that the $g^{(2)}(0)$ correlation function is directly transferred from the GW onto the detector, and it does not depend on the coupling strength $\gamma_\g$, regardless of the quantum state of the gravitational wave. We will comment on this behavior below. This is in contrast to other metrics of non-classicality, such as the Mandel $Q$ parameter, which in our case takes the form $Q_\br=\sin^2(\gamma_\g t)\langle \hat{n}_\gr\rangle \big(g_\gr^{(2)}(0)-1\big)$. These and other quantities depend manifestly on the quantity $\gamma_\g t$, whereas the full $g^{(2)}(0)$ does not.

\begin{figure}[t]
    \centering
    \includegraphics[width=1\linewidth]{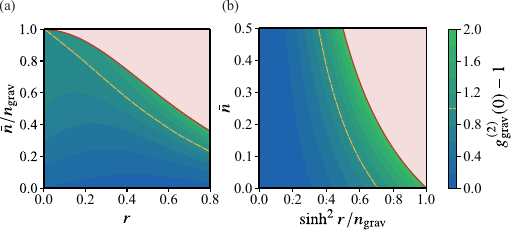}
    \caption{The $\g^{(2)}$-function of the GW is directly mapped on the detector, allowing for the characterization of the quantum statistics and state of the radiation. The figure shows contour plots of $\g^{(2)}_\gr(0)-1$ for gravitational wave signals with strain $h=10^{-22}$, $\theta=0$, with the total number of gravitons given by Eq.~\eqref{eq:average_graviton_count} with $n_\q \lesssim n_\gr$. 
    Fig.~(a) corresponds to the limit $\bar{n}\approx n_\q$, and Fig.~(b) corresponds to $\sinh^2r\approx n_\q$. In both panels, the yellow dashed curve corresponds to $g_\gr^{(2)}(0)=2$ as for an exactly thermal state. For Gaussian signals, this value is exceeded through squeezing in the gravitational wave.}
    \label{fig:g2examples}
\end{figure}

If the gravitational wave is in the general Gaussian state specified in Eq.~\eqref{eq:general_gaussian_gravitational_wave_state}, using Eq.~\eqref{eq:s_ordered_moments}, we find the general result
\begin{widetext}
\begin{align}
\label{eq:g2_general_gaussian_state}
    g_\br^{(2)}(0)=1+2\frac{2 \left(2 |\alpha|^2-1\right)\! \left(\bar{n}+\frac{1}{2}\right)\! \cosh (2 r)-4 |\alpha|^2 \!\left(\bar{n}+\frac{1}{2}\right) \sinh (2 r) \sin (2 \theta )-2 |\alpha|^2+2\!\left(\bar{n}+\frac{1}{2}\right)^2\! \cosh (4 r)+\frac{1}{2}}{\left[2 |\alpha|^2+2 \left(\bar{n}+\frac{1}{2}\right) \cosh (2 r)-1\right]^2}.
\end{align}
\end{widetext}    
This expression shows how a general Gaussian radiation state affects the counting statistics, resulting in distinct signatures that can be probed. Furthermore, rather than measuring the full energy fluctuations as in Eq.~\eqref{eq:general_bar_second_order_zero_time_delay}, one can instead link the second-order coherence \eqref{eq:second_order_autocorrelation_function}  directly to ratios of the excitation probabilities, as was proposed in Ref.~\cite{sreenath2025}. In fact, expanding to lowest order in $\gamma_\g t$ the probabilities $\mathbb{P}_0(t)$, $\mathbb{P}_1(t)$, and $\mathbb{P}_2(t)$, of detecting, respectively, zero, one, and two excitations in the bulk resonator calculated in the section above, yield
\begin{align}
\label{eq:g_2_as_ratio_test}
   \left. 2\frac{\mathbb{P}_0(t)\mathbb{P}_2(t)}{\mathbb{P}_1^2(t)}\right|_\br=g_\gr^{(2)}(0)+O\left((\gamma_\g t)^2\right).
\end{align}
These probabilities of detection are thus sufficient to determine the statistics of the GW radiation field. The left-hand side of Eq.~\eqref{eq:g_2_as_ratio_test} corresponds precisely to the ratio test $R$ defined in Ref.~\cite{sreenath2025}. The result in Eq.~\eqref{eq:g_2_as_ratio_test} is valid for any arbitrary GW states, provided that the interaction between the resonator and the GW is weak.  

The measurement of the second-order correlations \eqref{eq:general_bar_second_order_zero_time_delay}, either through direct phonon intensity-correlation measurements, or through the ratio \eqref{eq:g_2_as_ratio_test}, provides a direct means to determine the counting statistics of the passing GW through measurements of the phonons in the resonant detector, distinguishing the various possible states (and possible quantum signatures), as shown in Fig.~\ref{fig:g2examples}. In contrast to the direct comparison of the excitation probabilities for different states discussed in the previous section, the determination of the counting statistics can be achieved regardless of the weak coupling to the gravitational field. Beyond Gaussian states, since the $g_\gr^{(2)}(0)$-function is directly measured, potential sub-Poissonian statistics of the GW would also be directly detectable.

\subsection{Renormalization of the vacuum state second-order correlation function and thermal noise}
The apparent interaction-independence of the $g^{(2)}(0)$ function observed above is a consequence of the way in which the $g^{(2)}(0)$ function is normalized and the fact that the initial state of the detector mode is the vacuum. 
In fact, the second-order coherence as defined by Eq.~\eqref{eq:second_order_autocorrelation_function} is ill-defined in the case of a vacuum state, since $g^{(2)}(0)=0/0$. It is possible to attribute a limiting value if the vacuum state is defined as a limit of a general Gaussian state, but such a limit would not define a unique value for the correlation function. For example, we can equally view the vacuum state as a thermal state where the thermal occupation number approaches zero, that is, $\pureket{0}_\br=\lim_{n_\text{th}\to0}\rho_{n_\text{th}}$, or as a coherent state where $\alpha\to 0$, namely $\pureket{0}_\br=\lim_{\alpha\to0}\pureket{\alpha}$. In the first case, the vacuum limit yields the correlation function $g^{(2)}(0)=2$, while the latter yields $g^{(2)}(0)=1$. 

In light of this, the initial value of the second-order correlation function should be consistent with the state preparation process that brings the detector to its ground state. Physically, for the detectors envisioned here, this corresponds to the ground-state cooling of the mechanical mode of the bar resonator. To illustrate the last point, let us consider the case in which the detector is initially in a thermal state with thermal occupation number $n_\text{th}$ \footnote{Note that the resonator's average thermal occupation number $n_\text{th}$ should not be confused with $\bar{n}$, that is, the graviton thermal occupation number.}. In App.~\ref{app:second_order_correlation_appendix}, we determine the second-order correlation function for the case of general Gaussian GWs. For simplicity, let us consider $\bar{n}=0$, corresponding to the GW being in a displaced squeezed state. In this case, the second-order correlation function evaluates to
\begin{align}
\label{eq:g2_function_example_thermal_state_bar}
    g_{\br}^{(2)}(0)=2-\frac{\sin^4(\gamma_\g t) \left[|\alpha|^2-\frac{1}{2}\sinh(2r)\right]^2}{ \left[ n_\text{th} +\sin ^2(\gamma_\g t)(|\alpha| ^2+\sinh^2r- n_\text{th})\right]^2},
\end{align}
which depends explicitly on the interaction strength, and is well-defined for $t=0$. In particular, from Eq.~\eqref{eq:g2_function_example_thermal_state_bar} we can see that $1\leq g^{(2)}(0)\leq 3+(\sinh^2r)^{-1}$, for $t\geq0$. When $n_\text{th}\to0$, the $g_\br^{(2)}(0)$ defined exhibits a discontinuity of the first kind at $t=0$. But in all physically relevant cases, i.e., when the occupation number $n_\text{th}$ is finite but sufficiently small, the $g_\br^{(2)}(0)$ function is smoothly defined for all values of $t$. This suggests that deviations from the initial value of the detector's second-order coherence are detectable as long as $n_\text{th}\ll n_\gr (\gamma_\g t)^2$, which holds for general Gaussian GW signals---as can be seen from Eq.~\eqref{eq:second_order_correlation_noisy_detector}. In addition to the initial detector noise, we also need to avoid phonon gain during the measurement. As detailed in App.~\ref{app:second_order_correlation_appendix}, this corresponds to the requirement $\Gamma_\text{th} \,t \ll n_\gr (\gamma_\g t)^2$, where $\Gamma_\text{th}$ is the heating rate. In general, for a detector of quality factor $Q$, the latter condition is more stringent than the condition on $n_{\text{th}}$.

The discussion above aligns with the intuition that our ability to detect the correlations in the gravitational field through resonant detectors is as good as our ability to cool and maintain the detector in its ground state. In addition, to determine the correlation function \eqref{eq:second_order_autocorrelation_function}, one needs to measure the numerator and denominator separately, with the latter serving as the normalization factor. From there, to obtain a signal, each of these must be sufficiently large (which is why the detection can only work if the excitations in Eq.~\eqref{eq:excitations} are at least of order unity).

\section{State tomography through phase sensitive measurements}
\label{sec:state_tomography_through_phase_sensitive_measurements}
We now show that we can also go beyond counting statistics of the GWs and obtain full tomography of Gaussian states. When the detector is fixed to a reference state (as in our discussion so far, where it was assumed to be initially in the ground state), the second-order correlation function \eqref{eq:g2_general_gaussian_state} contains only partial tomographic information of the Gaussian signal. To completely determine the Gaussian state and ultimately achieve full tomography, it is necessary to access the phase of the incoming gravitational wave signal \cite{manikandan2025complementary,loughlin2025wave}. In other words, if we define the generalized gravitational wave quadrature operator
\begin{align}
    \hat{h}_{\phi}(t)=\frac{1}{\sqrt{2}}\left[e^{-i\phi}\hat{a}(t)+e^{i\phi}\hat{a}^\dagger(t)\right],
\end{align} 
our goal is to determine the value of the quadrature for different choices of $\phi$; this can be accomplished by preparing the state of the detector in a reference state with a known phase. By varying the reference phase and recording the corresponding response of the detector, different quadratures can be accessed.

\subsection{Homodyne intensity correlation measurement scheme}
For optical signals, the standard phase-sensitive measurement schemes are Homodyne or Heterodyne measurements. Here, we consider the so-called homodyne intensity correlation scheme \cite{vogel1995}. The strength of this method lies in the fact that, by using correlation measurements, detector inefficiencies only appear as multiplicative factors that don't spoil the signal \cite{vogel1995}. This is unlike standard homodyne measurement schemes. The scheme consists of measuring the intensity correlations of a signal that has been superimposed on a beamsplitter with a reference local oscillator. In our case, the signal is the GW, and the mixing of the signal to the local oscillator happens through the gravitational interaction with the bar detector. We consider then the unnormalized time-delayed second-order coherence of the detector, given by
\begin{align}
\label{eq:second_order_delayed_intensity_correlation}
    G^{(2)}_\br(\tau)=\langle\hat{b}^\dagger(t)\hat{b}^\dagger(t+\tau)\hat{b}(t+\tau)\hat{b}(t)\rangle.
\end{align}
Recalling the beamsplitter relations in Eq.~\eqref{eq:gravitational_beam_splitter}, and assuming that the resonator is in a coherent state of amplitude $\beta\in\mathbb{C}$, we see that the correlation function \eqref{eq:second_order_delayed_intensity_correlation} is composed of five terms each of order $|\beta|^k$, with $k\in \{0,\dots,4\}$. That is, $G^{(2)}(\tau)=\sum_{k=0}^{4} G^{(2)}_{\br,k}(\tau)$. As pointed out in Ref.~\cite{vogel1995}, not all of the contributions are detectable in correlation measurements. We are therefore interested in separating the different contributions. To that end, it is convenient to compare the short- and long-time delay correlations, since physical correlation functions tend to de-correlate for long time delays. Thus, we consider the quantity
\begin{align}
    \label{eq:delta_second_order_correlation}
    \Delta G^{(2)}_\br=G^{(2)}_\br(0)-\lim_{\tau\to\infty}G^{(2)}_\br(\tau).
\end{align}
Given the weakness of the coupling of the GW modes to the resonator, i.e., $\gamma_\g t\ll1$, we have $G^{(2)}_{\br,4}(0)\approx \lim_{\tau\to\infty}G^{(2)}_{\br,4}(\tau)$, which implies $\Delta G^{(2)}_{\br,4}\approx0$. We are then left with four contributions. In particular, we find the zeroth-order term
\begin{align}
\label{eq:zeroth_order_delta_non_static}
    \Delta G^{(2)}_{\br,0}=\sin^4(\gamma_\g t)\left[\langle:\!\hat{n}^2_\gr(t)\!:\rangle-\langle\hat{n}_\gr(t)\rangle\langle\hat{n}_\gr(\infty)\rangle\right],
\end{align}
the first-order term
\begin{align}
\begin{split}
    \Delta G^{(2)}_{\br,1}=&\sin^3(\gamma_\g t)|\beta|\cos(\gamma_\g t)\left[2\langle:\!\hat{h}_{\phi}(t)\hat{n}_\gr(t)\!:\rangle\right.\\&-\left.\langle\hat{h}_{\phi}(t)\rangle\langle\hat{n}_\gr(\infty)\rangle-\langle\hat{n}_{\gr}(t)\rangle\langle\hat{h}_\phi(\infty)\rangle\right],
\end{split}
\end{align}
the second-order term
\begin{align}
\begin{split}
    \Delta G^{(2)}_{\br,2}=&\sin^2(\gamma_\g t)|\beta|^2\cos^2(\gamma_\g t)\left[\langle:\!\hat{h}^2_\phi(t)\!:\rangle\right.\\&\left.-\langle\hat{h}_\phi(t)\rangle\langle\hat{h}_\phi(\infty)\rangle+\langle\hat{n}_\gr(t)\rangle-\langle\hat{n}_\gr(\infty)\rangle\right],
\end{split}
\end{align}
and, finally, the third-order contribution
\begin{align}
    \Delta G^{(2)}_{\br,3}=&\sin(\gamma_\g t)|\beta|^3\cos^3(\gamma_\g t)\left[\langle\hat{h}_\phi(t)\rangle-\langle\hat{h}_\phi(\infty)\rangle\right].
\end{align}
Gravitational waves are generally not stationary signals. However, we can assume stationarity of the signal around the resonance frequency in the detection window $\Delta t$. In that case, the only remaining contributions are
\begin{gather}
    \label{eq:zeroth_order_delta}
    \Delta G^{(2)}_{\br,0}=\sin^4(\gamma_\g t)\langle:\!(\Delta\hat{n}_\gr)^2\!:\rangle,
    \\
    \label{eq:first_order_delta}
    \Delta G^{(2)}_{\br,1}=\sin^3(\gamma_\g t)|\beta|\cos(\gamma_\g t)\langle:\!\Delta\hat{h}_{\phi}\,\Delta\hat{n}_\gr\!:\rangle,
    \\
    \label{eq:second_order_delta}
    \Delta G^{(2)}_{\br,2}=\sin^2(\gamma_\g t)\,|\beta|^2\cos^2(\gamma_\g t)\langle:\!(\Delta\hat{h}_{\phi})^2\!:\rangle,
\end{gather}
where we defined the deviation from the mean of an operator $\hat{A}$ as $\Delta\hat{A}=\hat{A}-\langle\hat{A}\rangle$, and $: \!\hat{\Pi}\!:$ stands for the normal-ordering of products $\hat{\Pi}$ of creation and annihilation operators. Now, supposing $\Delta G^{(2)}_{\br}$ has been measured, we need further criteria to separate the remaining terms in Eqs.~\eqref{eq:zeroth_order_delta}--\eqref{eq:second_order_delta}.

\subsection{Physical interpretation and separation of effects}
The zeroth-order contribution in Eq.~\eqref{eq:zeroth_order_delta}---or equivalently, Eq.~\eqref{eq:zeroth_order_delta_non_static} for the non-static case---corresponds to the normal-ordered intensity fluctuation correlations of the gravitational field. This term can be independently measured in the limit $|\beta|\to0$. In fact, up to normalization, Eq.~\eqref{eq:zeroth_order_delta} corresponds precisely to the zero-time delay second-order autocorrelation function in Eq.~\eqref{eq:general_bar_second_order_zero_time_delay} discussed in the previous section. 

The first-order term $\Delta G^{(2)}_{\br,1}$ in Eq.~\eqref{eq:first_order_delta} contains critical information about the GW state. It represents the normal-ordered correlation between the quadrature and intensity fluctuations of the field. This term exactly vanishes for many types of radiation, like Fock states, coherent states, squeezed states, or thermal states---see App.~\ref{app:field_correlations}. In contrast, it is non-vanishing for displaced squeezed and displaced thermal states, the states of interest here. It thus provides a means to discern a simple, coherent state GW input from a more general state, and is obtained from the measurement scheme described above. 

Finally, the second-order term $\Delta G^{(2)}_{\br,2}$ in Eq.~\eqref{eq:second_order_delta} contains the normal-ordered quadrature fluctuations of the gravitational field. For Gaussian states, the measurement of this term allows for full state tomography. To discriminate between this contribution and the others, one can notice the different periodicity with respect to the phase $\phi$, which enables one to isolate this term in a practical experiment, by varying the phonon state $\ket{\beta}$.  Thus, absorption measurements alone are sufficient to obtain full information about the passing GW state, exploiting the ability to prepare different input states of the bar and perform counting statistics at different times. 

We now discuss a concrete example. For GWs whose average number of non-coherent excitations $n_\q$ is comparable to the total average occupation number $n_\gr$, the three terms \eqref{eq:zeroth_order_delta}--\eqref{eq:second_order_delta} are all of order unity, and can therefore be detected in this measurement scheme. For example, for a displaced squeezed gravitational wave ($\bar{n}=0$) with the relative phase $\theta=0$, starting from the expressions derived in App.~\ref{app:field_correlations}, and considering the limit of large squeezing $r\gg1$, we find 
\begin{align}
\label{eq:first_order_limit}
    \Delta G^{(2)}_{\br,1}\approx\frac{1}{2} |\beta|\sin^3(\gamma_\g t)\cos(\phi)|\alpha|e^{2r},
\end{align}
for the first-order term, and
\begin{align}
\label{eq:second_order_limit}
    \Delta G^{(2)}_{\br,2}\approx\frac{1}{2} |\beta|^2\sin^2(\gamma_\g t)\sin^2(\phi)e^{2r},
\end{align}
for the second-order term. This shows, explicitly, the different phase dependence of the two terms. Furthermore, depending on the quadrature measured, i.e., on the choice of $\phi$, the term \eqref{eq:first_order_limit} can acquire negative values. In general, this could lead to an overall sub-Poisson statistic. However, in this particular case, the addition of the zeroth order term, which for $r\gg1$ is  $\Delta G^{(2)}_{\br,2} \simeq \sin^4(\gamma_\g t) e ^{4r}$, ensures that the overall $\Delta G^{(2)}_\br$ is consistent with super-Poissonian counting statistics, as should be the case for Gaussian states. For more exotic GW states (if these exist), this need not be the case, and the first-order term could contribute to signatures of sub-Poissonian statistics.

\subsection{Limits to quantum state tomography}
While we have shown above that quantum state tomography can be achieved for Gaussian GW signals, it will have practical limitations. In particular, there will be an experimental limit on how well $G^{(2)}(\tau)$ can be determined, due to imperfections, finite timing, detector noise, and other deleterious sources. This will limit the ability to determine the states of the GW. If we denote the imprecision by $\delta_{\textrm{exp}}=|G^{(2)}-G^{(2)}_\text{exp}|$, then our method can determine GW states with quadrature variances of order $\langle (\Delta h_{\phi})^2 \rangle \gtrsim \delta_{\text{exp}} /(\gamma_\g^2 t^2 \,|\beta|^2)$, where $\beta$ is the local oscillator amplitude, in this case, the initial state in the bar detector. Thus, in practice, the ability to discern arbitrary GW states will be limited, and it is best achieved with a strong acoustic drive and for long times, i.e., for continuous sources, such as ring-downs of neutron star mergers \cite{lasky2015gravitational, tobar2025detecting} or pulsars \cite{hulse1975discovery, singh2017detecting}. Recently, an acoustic superfluid resonator was excited to beyond $|\beta|^2 > 3 \times 10^4$ and its $g^{(2)}(0)$-function measured \cite{patil2022measuring}, demonstrating the feasibility of such experiments. 

However, as first noted in Ref.~\cite{ou1987}, in the homodyne correlation measurement scheme, one is also limited by classical noise coming from amplitude fluctuations in the local oscillator. This is because such noise is not balanced out like in traditional homodyne schemes. As a result, there is a trade-off between signal amplification on the one hand and noise on the other. To illustrate this, let us consider a simple analysis of noise originating from classical Gaussian amplitude fluctuations of the detector. We denote the fluctuations around $\beta$ by $\delta \beta(t)$ and assume that the time averages of the fluctuations satisfy \begin{align}
    \overline{\delta\beta(t)}=0, \quad\overline{\delta\beta(t)\beta(t+\tau)}=\overline{(\delta\beta)^2}e^{-\kappa \tau},
\end{align}
where $\tau>t>0$, $\kappa$ is the correlation decay timescale of the fluctuations, and $\overline{(\delta\beta)^2}$ is the variance of the fluctuations. Taking into account such classical noise, we can compute the second-order correlation \eqref{eq:second_order_delayed_intensity_correlation} averaged over the fluctuations. Notably, in taking the difference \eqref{eq:delta_second_order_correlation}, the leading order term $|\beta|^4$ no longer cancels, rather, we find \cite{vogel1995}
\begin{align}
\label{eq:fluctuation_fourth_order}
    \Delta\overline{G_{\br,4}^{(2)}}\approx 4\cos^4(\gamma_\g t) |\beta|^2\overline{(\delta \beta)^2},
\end{align}
where we assumed that $\overline{(\delta \beta)^2}/|\beta|^2=\varepsilon\ll1 $. If we are interested in the measurement of the quadrature variance from Eq.~\eqref{eq:second_order_delta}, by comparing with the leading order fluctuation term \eqref{eq:fluctuation_fourth_order}, we get the signal-to-noise ratio (SNR)
\begin{align}
\label{eq:snr}
    \left|\frac{\Delta G_{\br,2}^{(2)}}{\Delta\overline{G_{\br,4}^{(2)}}}\right|=\frac{\sin^2(\gamma_\g t)\langle:\!(\Delta\hat{h}_{\phi})^2\!:\rangle}{4\,\varepsilon |\beta|^2}.
\end{align}
From Eq.~\eqref{eq:snr}, we see that if the amplitude of the detector is of the same order of magnitude as the quadrature variance of the signal, namely, if $|\beta|^2\sim\sin^2(\gamma_\g t)\langle:\!\!(\Delta\hat{h}_{\phi})^2\!\!:\rangle$,  the ratio \eqref{eq:snr} becomes of order $1/\varepsilon\gg1$. This shows the trade-off explicitly; by increasing the amplitude $|\beta|$, the signal in Eq.~\eqref{eq:second_order_delta} is boosted, but at the same time the SNR decreases. Nonetheless, the relation Eq.~\eqref{eq:snr} also shows that the SNR can remain large for larger amplitudes $|\beta|$, if the ratio $\varepsilon$ is sufficiently small, giving leeway to boost weaker GW signals.

\subsection{Weak coupling not a limitation} 
A key feature of the described approach to tomography is that access to the second-order correlations of the GW is enabled through intensity-correlations measurements, even when the graviton–matter coupling is extremely weak and detector efficiencies are very low. The use of correlation measurements for state reconstruction is well established in quantum optics due to their inherent robustness against detector inefficiencies \cite{ou1987,vogel1995,schulte2015quadrature}. We find that the same rationale applies to gravitational waves, making the detection of gravitational-wave statistics possible despite the weakness of the interaction. 

Beyond the close analogy with quantum-optical detection methods, important differences arise in practical implementation. In standard optical homodyne detection, the local oscillator can typically be made arbitrarily strong, and excess noise can be suppressed through careful power balancing, leading to precise quadrature measurements. In contrast, the effective local oscillator in our scheme is realized by a driven mechanical mode of the detector, whose amplitude cannot be increased arbitrarily without introducing excess amplitude noise. As discussed above, this noise limits the achievable signal-to-noise ratio by introducing additional spurious correlations that cannot be balanced out. A further experimental challenge is the control of the phonon-mode phase, which is required to perform full Gaussian state tomography. We also highlight that the method requires measurements over a long time-period, meaning that either several detectors have to be used, or that continuous GW sources, such as pulsars \cite{singh2017detecting}, are desirable.

These considerations imply that progress in detecting the quantum statistics of gravitational waves, through the methods described in this manuscript, will be driven primarily by improvements in detector coherence, thermal occupation, and classical noise suppression, circumventing the weak graviton–to-matter coupling and without the need for near-unit detector efficiency.

\section{Conclusions}
Our work establishes a bridge between quantum optics and gravitational-wave physics at the level of quantum measurement theory. We show that graviton detection can serve as a practical tool for extracting the quantum statistics of gravitational waves and determining their state as well as their potential quantum features. Graviton detection was only recently shown to become experimentally accessible in the near future \cite{tobar2024} due to rapid advances in quantum acoustics of macroscopic resonators, and the ability to extract single quanta of energy from passing waves. Here, we derive the graviton-to-phonon conversion probabilities for general Gaussian gravitational waves and show that these probabilities allow for the discrimination of different Gaussian radiation states. In addition, we demonstrate that the second-order correlation function of the gravitational wave is directly transferred to the detector, irrespective of the weak gravitational coupling, enabling direct measurement of GW counting statistics. Building on this result, we present a scheme based on coherent acoustic driving with known reference phases that enables full tomography of Gaussian gravitational radiation states. Thus, in addition to tests of linearized quantum gravity \cite{shenderov2024stimulated}, graviton counting also opens new insights into the composition and sources of gravitational waves, and can potentially reveal non-classical features of the radiation. By showing how standard measurement tools of quantum optics can be adapted to the gravitational context, this work opens a pathway toward characterizing the quantum states of gravitational radiation and exploring quantum aspects of gravity using near-future quantum sensing technologies.

\begin{acknowledgments}
{We thank Jonatan Bohr Brask, Konstantin Beyer, and Jack Harris for their valuable insights and comments. This work was supported by the National Science Foundation under Grant No.\ 2239498 and by NASA under Grant No.\ 80NSSC25K7051. KT and ULA gratefully acknowledge support from the Danish National Research Foundation, Center for Macroscopic Quantum States (bigQ, DNRF142), and Novo Nordisk Foundation (CBQS NNF 24SA0088433). KT thanks the Otto Mønsted Foundation, Augustinus Foundation, and William Demant Foundation for supporting the research stay at Stevens Institute of Technology during which this work was developed.}
\end{acknowledgments}

\bibliography{quantum_state_characterization_of_gravitational_waves_via_graviton_counting_statistics}

\appendix
\onecolumngrid
\section{Interaction Hamiltonian}
\label{app:interaction_hamiltonian_derivation}
We consider incident gravitational radiation on a macroscopic bar resonator. We describe a resonator of mass $M$ as a one-dimensional chain of $N+1$ (with $N$ an odd positive integer) interacting atoms arranged along a line of length $L$. We restrict to nearest neighbor interaction. Each atom will oscillate around its center of mass $x_n=a n/2$, where $a$ is the lattice spacing and $-N<n<N$ is an odd integer. The equations of motion for a displacement $\xi_n$ of the $n$-th atom are easily solved, leading to 
\begin{align}
    \xi_n(t)=e^{-i\omega t}\left(Ae^{ikna/2}+Be^{-ikna/2}\right)+\text{H.c.},
\end{align}
with $\omega^2=2\,\omega_D^2(1-\cos(ka))$, $m\,\omega^{}_D$ the tension per unit length of the resonator, and $m=M/N$. The classical Hamiltonian describing the interaction is given by 
\begin{align}
    H_\text{int}=-\frac{ML \ddot{h}_{xx}}{\pi^2}\sum_{\ell=1,3,\dots}^{N} (-1)^\frac{\ell-1}{2} \frac{1}{\ell^2}\chi^{}_\ell(t)-\frac{M \ddot{h}_{xx}}{8}\sum_{\ell=0,2,\dots}^{N-1}\chi^{2}_\ell(t).
\end{align}
The dominant terms in the interaction arise from the odd series. Upon quantization, the oscillator modes become operators, namely $\chi^{}_\ell\rightarrow\hat\chi^{}_\ell=\sqrt{\hbar/2m_\ell\omega_\ell}\left(\hat{b}_\ell+\hat{b}^\dagger_\ell\right)$, and similarly the flat metric perturbation due to gravity becomes an operator. Assuming the single-mode and dipole approximations, we find that for odd-$\ell$
\begin{align}
     H_\text{int}=\hbar\, \gamma_\g\left(\hat{b}_\ell+\hat{b}^\dagger_\ell\right)\left(\hat{a}\, e^{-i \nu t}+\hat{a}^\dagger e^{i \nu t} \right),
\end{align}
where $\nu=\nu(t)$ is the gravitational wave frequency,
\begin{align}
    \gamma_\g=\sqrt{(-1)^{\ell-1}\frac{8\pi G M\nu^3L^3}{\omega_\ell\, c^2 V \pi^4 \ell^4}},
\end{align} 
$M$ is the mass of the bar, $L$ is the bar length, and $V$ is the characteristic volume of the gravitational wave.

\section{Heisenberg evolution}
\subsection{Evolution in the rotating wave approximation}
\label{app:evolution_in_the_rotating_wave_approximation}
If we perform the rotating wave approximation, we can write the full Hamiltonian as
\begin{align}
    H_\ell=\hbar\omega_\ell \, \hat{b}^\dagger_\ell \hat{b}_\ell+\hbar\, \gamma_\g\left(\hat{b}_\ell\,\hat{a}^\dagger e^{i \nu t}+\hat{b}^\dagger_\ell\, \hat{a}\, e^{-i \nu t}\right).
\end{align}
We can very easily solve the dynamics in the Heisenberg picture. We are interested in determining the evolution of the operators
\begin{align}
    \frac{\dd }{\dd t}\hat{b}_\ell = \frac{i}{\hbar} \left[\hat{H},\hat{b}_\ell\right],\quad \frac{\dd }{\dd t}\hat{a} = \frac{i}{\hbar} \left[\hat{H},\hat{a}\right] +\frac{\partial \,\hat{a}}{\partial \,t}.
\end{align}
We see that this determines a system of coupled differential equations
\begin{align}
\begin{cases}
   \dot{\hat{b}}_\ell = -i \left(\omega_\ell \,\hat{b}_\ell + \gamma_\g\, \hat{a} \right) \\ \dot{\hat{a}} = -i \left(\nu \,\hat{a}+ \gamma_\g\, \hat{b}_\ell \right) .
\end{cases}
\end{align}
The general solution to this system is of the form
\begin{align}
        \hat{b}_\ell(t) = e^{-i\left(\omega_\ell+\nu\right) t/2} \left[\hat{a}(0) f(t)+\hat{b}(0) g_-(t)\right], \quad \hat{a}(t) = e^{-i\left(\omega_\ell+\nu\right) t/2} \left[\hat{b}(0) f(t)+\hat{a}(0) g_+(t)\right],
\end{align}
where
\begin{gather}
    f(t)=e^{- i t \lambda_\g/2}\frac{ \gamma_\g  \left(1-e^{i t \lambda_\g}\right)}{\lambda_\g}, \quad
    g_\pm(t)=e^{- i t \lambda_\g/2}\frac{\lambda_\g+\left(\lambda_\g-\Delta \right) e^{i t \lambda_\g}\pm\Delta}{2 \lambda_\g},
\end{gather}
and where we defined $\lambda_\g=\sqrt{4 \gamma_\g ^2+\Delta^2}$, and the detuning $\Delta= \nu - \omega_\ell$. At resonance, i.e., when $\Delta=0$, we have
\begin{align}
        \hat{b}_\ell(t) = e^{-i \omega_\ell t} \left[ \cos(\gamma_\g t)\,\hat{b}_\ell(0) -i \sin(\gamma_\g t)\,\hat{a}(0)  \right],\,\,
        \hat{a}(t) =  e^{-i \omega_\ell t}\left[\cos(\gamma_\g t)\,\hat{a}(0) -i \sin(\gamma_\g t)\,\hat{b}_\ell(0)  \right].
\end{align}
Having solved the equations of motion for the ladder operators, we can compute the time evolution of any observable we might be interested in. 

\subsection{Beyond the rotating wave approximation}
\label{app:evolution_beyond_the_rotating_wave_approximation}
For completeness, we discuss here the solution to the dynamics \eqref{eq:interaction_hamiltonian} in the resonant case ($\Delta=0$) without assuming the rotation wave approximation. In this case, the equations of motion are given by the coupled equations 
\begin{align}
\begin{cases}
   \dot{b}_\ell = -i \omega_\ell \,b_\ell -i \gamma_\g\, (a +a^\dagger) \\
   \dot{a} = -i \nu \,a -i \gamma_\g\, (b_\ell+b^\dagger_\ell)  \\
   \dot{b}^\dagger_\ell = i \omega_\ell \,b^\dagger_\ell + i\gamma_\g\, (a+a^\dagger )\\
   \dot{a}^\dagger = i \nu \,a^\dagger +i \gamma_\g\,(b_\ell+ b^\dagger_\ell ).
\end{cases}
\end{align}
The solutions can be written in the form
\begin{align}
\begin{split}
    \hat{a}(t)=&\frac{1}{2}\left\{\hat{a}(0) \left[ \cos\left(\Omega_{\g}^{+}t\right)+\cos\left(\Omega_{\g}^{-}t\right)+\mathcal{A}(t)+\mathcal{B}(t)\right]+\hat{a}^\dagger(0)\,\mathcal{D}(t)\right.\\
    &\left.+\hat{b}(0)\,\left[ \cos\left(\Omega_{\g}^{+}t\right)-\cos\left(\Omega_{\g}^{-}t\right)+\mathcal{D}(t)+\mathcal{E}(t)\right]+\hat{b}^\dagger(0)\,\mathcal{A}(t)\right\},
\end{split}
\end{align}
where
\begin{align}
\begin{gathered}
    \mathcal{A}(t)=i\frac{ \gamma_\g}{\omega_\g}\left[\frac{\Omega_\g^+}{\omega}\sin \left(\Omega_\g^-
    \,t \right)- \frac{\Omega_\g^-}{\omega}
    \sin \left(\Omega_\g^+ \,t \right)\right],\quad
    \mathcal{B}(t)=-\frac{i}{\omega_\g}\left[\Omega_\g^+\sin \left(\Omega_\g^-
    \, t \right) + \Omega_\g^-
    \sin \left(\Omega_\g^+ \, t \right)\right], \\
    \mathcal{D}(t)=-i\frac{ \gamma_\g}{\omega_\g}\left[\frac{\Omega_\g^+}{\omega} \sin \left(\Omega_\g^-
    \,t \right) + \frac{\Omega_\g^-}{\omega}
    \sin \left(\Omega_\g^+ \,t \right)\right], \quad
    \mathcal{E}(t)= \frac{i}{\omega_\g}\left[\Omega_\g^+ \sin \left(\Omega_\g^-
    \, t \right) - \Omega_\g^-
    \sin \left(\Omega_\g^+ \, t \right)\right],
\end{gathered}
\end{align}
with $\delta_\g=2 \gamma_\g/\omega$, $\Omega_\g^\pm=\omega\sqrt{1\pm\delta_\g}$ and $\omega_\g=\omega\sqrt{1-{\delta_\g}^2}$. Similarly, we find
\begin{align}
\begin{split}
    \hat{a}(t)=&\frac{1}{2}\left\{\hat{a}(0) \left[ \cos\left(\Omega_{\g}^{+}t\right)-\cos\left(\Omega_{\g}^{-}t\right)+\mathcal{D}(t)+\mathcal{E}(t)\right]+\hat{a}^\dagger(0)\,\mathcal{D}(t)\right.\\
    &\left.+\hat{b}(0)\,\left[ \cos\left(\Omega_{\g}^{+}t\right)+\cos\left(\Omega_{\g}^{-}t\right)+\mathcal{A}(t)+\mathcal{B}(t)\right]+\hat{b}^\dagger(0)\,\mathcal{A}(t)\right\}.
\end{split}
\end{align}

\subsection{Warm-up: transfer of squeezing from a squeezed vacuum gravitational wave signal}
To build an intuitive understanding of the dynamics determined by the Hamiltonian \eqref{eq:interaction_hamiltonian_rwa} in the rotating wave approximation, we consider the case where the incident gravitational wave is a squeezed vacuum state, and the local resonator (the detector) is in its ground state, that is, $|\Psi\rangle = |\xi\rangle_\text{grav} \otimes |0\rangle_\text{bar}$. Here, $|\xi \rangle=\hat{S}_\xi|0\rangle$ is the single-mode squeezed vacuum state, with squeezing $\xi=re^{i\varphi}$, which in the Fock basis $\{|n\rangle\}_{n=0}^\infty$ can be written as
\begin{align}
    |\xi\rangle =\frac{1}{\sqrt{\cosh(r)}} \sum_{n=0}^{\infty}\left(e^{-i\varphi}\tanh r\right)^n\frac{\sqrt{(2n)!}}{2^n n!}|2n\rangle.
\end{align}
 Let us consider the dynamics in the resonant case, i.e., $\Delta=0$. Having solved the Heisenberg equations of motion \eqref{eq:gravitational_beam_splitter}, we know the GW's interaction with the local oscillator is a beam-splitter interaction. This means that any squeezing present in the gravitational wave state should, in principle, be transferred to the detector on a timescale dictated by the coupling $\gamma_\g$. Let us try to estimate the magnitude of the squeezing transfer.

An arbitrary quadrature $\hat{q}$ is said to be squeezed if $\Delta\hat{q}(t)<\Delta\hat{q}_\text{zp}$, where $\Delta\hat{q}_\text{zp}$ is the zero point variance. Let us define the rotated quadrature
\begin{align}
    \hat q(\vartheta) = \hat{x}\cos(\vartheta)  + \hat{p} \sin(\vartheta)
\end{align}
where
\begin{align}
    \hat{x}=\frac{1}{\sqrt{2}}\left(\hat{b}^\dagger+\hat{b}\right)\,, \quad \text{and}\quad \hat{p}=\frac{i}{\sqrt{2}}\left(\hat{b}^\dagger-\hat{b}\right).
\end{align}
We have dropped the $\ell$ index since we are in the resonant case $\Delta=0$, and we have normalized each quadrature so that the zero point expectation values are $\Delta x_\text{zp}=\Delta p _\text{zp}=\Delta q_\text{zp}=1/\sqrt{2}$. Given that the oscillator state is the vacuum state and the gravitational wave is a squeezed vacuum state, $\langle q(\vartheta)\rangle=0$, and we have $\Delta q(\vartheta)=\sqrt{\langle q(\vartheta)^2\rangle}$. The expectation value we are after is given by
\begin{align}
    \langle q(\vartheta)^2 \rangle =\langle x^2 \rangle_t \cos^2(\vartheta) + \langle p^2 \rangle_t  \sin^2(\vartheta) + \langle \{x, p\}\rangle_t  \cos(\vartheta) \sin(\vartheta),
\end{align}
where $\{a,b\}=ab+ba$. Minimizing the expression above with respect to $\vartheta$, we find that the maximum amount of squeezing transferred to the Weber bar resonator at a given time is 
\begin{align}
    \langle q(\vartheta_\text{min})^2 \rangle = \frac{\langle x^2 \rangle_t +\langle p^2 \rangle_t }{2} -\frac{\sqrt{(\langle x^2 \rangle_t \!-\!\langle p^2 \rangle_t)^2+(\langle x p\rangle_t \!+\! \langle p x\rangle_t)^2}}{2},
\end{align}
where $\vartheta_\text{min}$ corresponds to the angle
\begin{align}
    \vartheta_\text{min}=\tan ^{-1}\left[\!\frac{\langle x p\rangle + \langle p x\rangle}{\langle x^2 \rangle \!- \!\langle p^2 \rangle\!-\!\sqrt{(\langle x^2 \rangle -\langle p^2 \rangle)^2+(\langle x p\rangle + \langle p x\rangle)^2}}\!\right].
\end{align}
Therefore, using the time-dependent operators derived in the appendix \ref{app:evolution_in_the_rotating_wave_approximation}, we find that
\begin{align}
    \langle q(\vartheta_\text{min})^2\rangle=\frac{1}{2}\left[1+\sin ^2(\gamma_\g  t)\left(e^{-2 r}-1\right) \right].
\end{align}
As expected, for $t=0$, no squeezing is present in the resonator. As $t$ increases, so does the squeezing of the quadrature $\hat{q}(\vartheta_\text{min})$, until it reaches its maximum at $t_\text{swap}=\pi/(2\gamma_\g)$ for which all the squeezing present in the gravitational wave is transferred to the resonator, i.e., $\Delta q(\vartheta_\text{min})=e^{-r}/\sqrt{2}$. Given the very small value of the coupling $\gamma_\g$, the timescale $t_\text{swap}$ is several times the age of the universe. Due to the Heisenberg uncertainty relations, the quadrature $\pi$-rotated with respect to $\vartheta_\text{min}$ will be maximally anti-squeezed, namely
\begin{align}
    \langle q(\vartheta_\text{max})^2\rangle=\frac{1}{2}\left[1+\sin ^2(\gamma_\g  t)\left(e^{2 r}-1\right) \right],
\end{align}
where $\vartheta_\text{max}=\vartheta_\text{min}+\pi$. For gravitational waves detectable by LIGO, the quantity $\sin^2(\gamma_\g t) e^{2r}$ can be of order unity, given the very high graviton occupation number $\langle \hat{n}_\gr\rangle\simeq10^{35}$---see Eq.~\eqref{eq:intensity}.

\section{Unitary evolution of Gaussian moments}
\label{app:gaussian_evolution}
For Hamiltonians quadratic in the canonical operators, like the interaction Hamiltonian \eqref{eq:interaction_hamiltonian} discussed in this manuscript, all initial  Gaussian states evolve into Gaussian states. A Gaussian state of $N$ modes is fully specified by the second and first-order moments, defined respectively by
\begin{align}
    \sigma_{ij}=\frac{1}{2}\langle\{\hat{r}_i,\hat{r}_j\}\rangle-\langle \hat{r}_i\rangle\langle\hat{r}_j\rangle,\,\text{and}\quad \bar{r}_i=\langle \hat{r}_i\rangle,
\end{align}
where $\hat{\mathbf{r}}=(\hat{x}_1,\hat{p}_1,\dots,\hat{x}_N,\hat{p}_N)^\text{T}$. For a general quadratic Hamiltonian, we can define the Hamiltonian matrix $H$ as
\begin{align}
    \hat{H}=\frac{1}{2}\hat{\boldsymbol{r}}^\text{T}H\hat{\boldsymbol{r}}+\hat{\boldsymbol{r}}^\text{T}\bar{\boldsymbol{r}},
\end{align}
and the evolution of an initial Gaussian state specified by $\boldsymbol{\sigma}_0$ and $\bar{\boldsymbol{r}}_0$, is given by the transformations
\begin{align}
    \boldsymbol{\sigma}(t)=S\boldsymbol{\sigma}_0 S^\text{T}, \quad \bar{\boldsymbol{r}}(t)=S\bar{\boldsymbol{r}}_0,
\end{align}
where $S=e^{\boldsymbol{\Omega}  Ht/\hbar}$, and $\boldsymbol{\Omega}$ is the $N$-dimensional symplectic form.  It is easy to see that $S\,\boldsymbol{\Omega}\, S^\text{T}=\boldsymbol{\Omega}$, meaning that $S$ is a symplectic transformation. 
\subsection{Evolution of two modes interacting through a beamsplitter Hamiltonian}
Let us employ the symplectic transformations defined above to determine the evolution of the Gaussian moments under the quadratic Hamiltonian
\begin{align}
 \hat{H}_\text{int}=\hbar\, \gamma_\g\left(\hat{a}\,\hat{b}^\dagger+\hat{a}^\dagger\, \hat{b} \right).
\end{align}
The symplectic transformation $S$ corresponding to this interaction Hamiltonian is given by 
\begin{align}   
\label{eq:gravitational_beam_splitter_transformation}
S= e^{\boldsymbol{\Omega}Ht/\hbar}=\begin{pmatrix}
     \cos(\gamma_\g t)& 0 & 0 & \sin(\gamma_\g t)\\
     \,0 & \cos(\gamma_\g t) & -\sin(\gamma_\g t) & 0\\
     0 & \sin(\gamma_\g t) & \cos(\gamma_\g t) & 0\,\\
     -\sin(\gamma_\g t) & 0 & 0 & \cos(\gamma_\g t)
\end{pmatrix},\,\text{where}\quad
H=\hbar\,\gamma_\g\begin{pmatrix}
     \,0& 0 & 1 & 0\,\\
     \,0 & 0 & 0 & 1\,\\
     \,1 & 0 & 0 & 0\,\\
     \,0 & 1 & 0 & 0\,
\end{pmatrix}.
\end{align}
 The most general separable initial state of the two interacting modes is given by
 \begin{align}
 \label{eq:general_initial_state}
\boldsymbol{\sigma}=    
\begin{pmatrix}
\boldsymbol{\sigma}_\gr&\boldsymbol{0}\\
\boldsymbol{0}& \boldsymbol{\sigma}_\br\\ 
\end{pmatrix}, \quad \bar{\boldsymbol{r}}=
\begin{pmatrix}
\bar{\boldsymbol{r}}_\gr\\
\bar{\boldsymbol{r}}_\br
\end{pmatrix},
\end{align}
with 
\begin{align}
\boldsymbol{\sigma}_i=    \begin{pmatrix}
     \,\left(\bar{n}_i+\frac{1}{2}\right) \left[\cosh (2 r_i)-\cos \theta_i\sinh (2 r_i)\right]& - \left(\bar{n}_i+\frac{1}{2}\right) \sin \theta_i\sinh(2 r_i)\,\\
     \,- \left(\bar{n}_i+\frac{1}{2}\right) \sin \theta_i\sinh(2 r_i)& \left(\bar{n}_i+\frac{1}{2}\right) \left[\cosh (2 r_i)+\cos \theta_i\sinh (2 r_i)\right]\,\\ 
\end{pmatrix}, \quad \bar{\boldsymbol{r}}_i=\sqrt{2}
\begin{pmatrix}
\text{Re}(\alpha_i)\\
\text{Im}(\alpha_i)
\end{pmatrix},
\end{align}
where $r_i$ is the squeezing parameter, $\theta_i$ the squeezing angle, $\bar{n}_i$ the average thermal occupation number, $\alpha_i$ the displacement, $i\in\{\gr,\br\}$, and  $\boldsymbol{0}$ the $2\times2$ null matrix. Applying the symplectic transformations \eqref{eq:symplectic_evolution} we straightforwardly find the evolved first and second moments. For example, the reduced evolved state of the second mode is given in terms of the covariance matrix
\begin{align}
\label{eq:evolution_general_covariance_matrix}
\boldsymbol{\sigma}_\br(t)=\begin{pmatrix}
    u_\br^{(+)}\cos^2 (\gamma_\g t)+u_\gr^{(-)} \sin^2 (\gamma_\g t) & v_\br \cos^2 (\gamma_\g t)-v_\gr \sin^2 (\gamma_\g t) \\
 v_\br \cos^2 (\gamma_\g t)-v_\gr \sin^2 (\gamma_\g t) & u_\gr^{(+)} \sin^2 (\gamma_\g t)+u_\br^{(-)} \cos^2 (\gamma_\g t) \\
\end{pmatrix},
\end{align}
and displacement vector
\begin{align}   
\label{eq:evolution_general_displacement_vector}\bar{\boldsymbol{r}}_\br(t)=\sqrt{2}\begin{pmatrix}
    \text{Im}(\alpha_\gr) \sin (\gamma_\g t)+\text{Re}(\alpha_\br) \cos (\gamma_\g t)\\
     \text{Im}(\alpha_\br) \cos (\gamma_\g t)-\text{Re}(\alpha_\gr) \sin (\gamma_\g t)
\end{pmatrix},
\end{align}
where 
\begin{align}
    u_i^{(\pm)}=\left[\cosh (2 s_i)\pm\cos \theta_i\sinh (2 s_i)\right]\left(\bar{n}_i+\frac{1}{2}\right) , \quad v_i=-  \sin \theta_i\sinh(2 s_i)\left(\bar{n}_i+\frac{1}{2}\right).
\end{align}

For example, if the second mode is initially in the vacuum state, we find that the reduced state of the second mode at time $t$ is specified by
\begin{align}
\boldsymbol{\sigma}_\br(t)=\sin^2(\gamma_\g t)\begin{pmatrix}
    \left(\frac{1}{2} +\bar{n}\right) \left[\cos (\theta ) \sinh (2 r)+\cosh (2 r)\right] & (2 \bar{n}+1) \sin (\theta ) \sinh (r) \cosh (r)  \\
    (2 \bar{n}+1) \sin (\theta ) \sinh (r) \cosh (r)& \left(\frac{1}{2} +\bar{n}\right)  \left[-\cos (\theta ) \sinh (2 r)+\cosh (2 r)\right] \\ 
\end{pmatrix}+\cos^2(\gamma_\g t) \frac{\id}{2},
\end{align}
and the displacement vector $\bar{\boldsymbol{r}}_\br(t)=\sin(\gamma_\g t) \left(\text{Im}(\alpha),-\text{Re}(\alpha)\right)^\text{T}$. We note that by applying the local phase shift 
\begin{align}
    R_{\frac{\pi}{2}}=\begin{pmatrix}
        0 & -1 \\
        1 & 0
    \end{pmatrix},
\end{align}
we can write the state of the second subsystem as
\begin{align}
\label{eq:evolved_covariance_and_displacement}
\begin{gathered}
    \boldsymbol{\sigma}_\br(t)= \cos^2(\gamma_\g t) \boldsymbol{\sigma}_\br(0)+\sin^2(\gamma_\g t)\mathcal{}\boldsymbol{\sigma}_\gr(0),\\ \bar{\boldsymbol{r}}_\br(t)=\sin(\gamma_\g t)\,\bar{\boldsymbol{r}}_\gr(0).     
\end{gathered}
\end{align}

\subsection{Open diffusive dynamics of the detector mode}
Above, we considered the unitary evolution of the joint gravitational wave and detector system. In practice, however, the resonant detector will be subject to environmental noise. Here, we consider a simple model based on the coupling of the detector to a large thermal bath under the assumption that the system-to-bath correlation time is shorter than the dynamical time scale (Markov assumption). This can be modeled through the Markovian evolution given by the equations 
\begin{align}
\label{eq:open_dynamics}
\begin{gathered}
    \frac{\dd}{\dd t}\boldsymbol{\sigma}=A \boldsymbol{\sigma} + \boldsymbol{\sigma}A^\text{T}+D, \\
    \frac{\dd}{\dd t}\bar{\boldsymbol{r}}= A \bar{\boldsymbol{r}} + \boldsymbol{d},
\end{gathered}
\end{align}
where $A=\boldsymbol{\Omega}H-\kappa\,\id/2$ is the drift matrix, $\kappa$ is a decay rate, $D=\kappa(\bar{N}+1/2)\id$, and  $\bar{N}$ the average number of thermal excitations in the environment \cite{serafini2023quantum}. 
Assuming the initial state is given by Eq.~\eqref{eq:general_initial_state} with the detector mode in the vacuum, the solutions to Eq.~\eqref{eq:open_dynamics} corresponding to the evolution of the bar mode is given by
\begin{align}
\label{eq:open_evolution_bar}
\begin{gathered}
    \boldsymbol{\sigma}_\br(t)=e^{-\kappa t}\left[ \cos^2(\gamma_\g t) \boldsymbol{\sigma}_\br(0)+\sin^2(\gamma_\g t)\boldsymbol{\sigma}_\gr(0)\right]+\left(1-e^{-\kappa t}\right) \left(\bar{N}+\frac{1}{2}\right)\id,\\ \bar{\boldsymbol{r}}_\br(t)=e^{-\kappa t/2}\sin(\gamma_\g t)\,\bar{\boldsymbol{r}}_\gr(0).     
\end{gathered}
\end{align}

\section{Explicit calculation of detection probabilities for general Gaussian states}
In this section, we explicitly show how to determine the detection probabilities discussed in the main text, starting from Eq.~\eqref{eq:loop_hafnian_detection_probabilities}. Starting from the covariance matrix and displacement vector defined in Eq.~\eqref{eq:gaussian_state_ladder_operator}, we can determine 
\begin{align}
    \boldsymbol{A}_\br=\begin{pmatrix}
        \mathfrak{a}_2 & \mathfrak{a}_1 \\
        \mathfrak{a}_1 & \mathfrak{a}_2
    \end{pmatrix}, \quad\boldsymbol{F}_\br=\begin{pmatrix}
        \mathfrak{f}_1\\
        \mathfrak{f}_2
    \end{pmatrix},
\end{align}
where
\begin{gather}
    \label{eq:def_a_1}
    \mathfrak{a}_1(t)=1-\frac{ \left(\frac{1}{2}+ \bar{n}\right) \cosh (2 r) \sin ^2(\gamma_\g t)+\frac{1}{2}\left(\cos ^2(\gamma_\g t)+2\right)}{\left(\frac{1}{2}+ \bar{n}\right) \cosh (2 r) \sin ^2(\gamma_\g t) (\cos^2(\gamma_\g t)+2)+\left(\frac{1}{2}+ \bar{n}\right)^2 \sin ^4(\gamma_\g t)+\frac{1}{4}\left(\cos ^2(\gamma_\g t)+2\right)^2} \\
    \label{eq:def_a_2}
    \mathfrak{a}_2(t)=-\frac{ \left(\frac{1}{2}+ \bar{n}\right) \sinh (2 r) \sin ^2(\gamma_\g t)}{\left(\frac{1}{2}+ \bar{n}\right) \cosh (2 r) \sin ^2(\gamma_\g t) (\cos^2(\gamma_\g t)+2)+\left(\frac{1}{2}+ \bar{n}\right)^2 \sin ^4(\gamma_\g t)+\frac{1}{4}\left(\cos ^2(\gamma_\g t)+2\right)^2} \\
    \label{eq:def_f_1}
    \mathfrak{f}_1(t)= \frac{8 |\alpha|  e^{-i \theta } \sin(\gamma_\g t) \left[(2 \bar{n}+1) \sin ^2(\gamma_\g t) \left(\sinh (2 r)+e^{i \theta } \cosh (2 r)\right)+e^{i \theta } \left(\cos ^2(\gamma_\g t)+2\right)\right]}{-8 \left(\bar{n}^2+\bar{n}-1\right) \cos (2\gamma_\g t)+4 (2 \bar{n}+1) \cosh (2 r) \sin ^2(\gamma_\g t) (\cos(2\gamma_\g t)+5)+(2 \bar{n} (\bar{n}+1)+1) \cos (4 \gamma_\g t)+6 \bar{n} (\bar{n}+1)+27} \\
    \label{eq:def_f_2}
    \mathfrak{f}_2(t)=\frac{8 |\alpha|  \sin(\gamma_\g t) \left[(2 \bar{n}+1) \sin ^2(\gamma_\g t) \left(\cosh (2 r)+e^{i \theta } \sinh (2 r)\right)+\cos ^2(\gamma_\g t)+2\right]}{-8 \left(\bar{n}^2+\bar{n}-1\right) \cos (2 \gamma_\g t)+4 (2 \bar{n}+1) \cosh (2 r) \sin ^2(\gamma_\g t) (\cos (2\gamma_\g t)+5)+(2 \bar{n} (\bar{n}+1)+1) \cos(4\gamma_\g t)+6 \bar{n} (\bar{n}+1)+27} .
\end{gather}
For the probabilities with $n=0,1,2$, we then have,
\begin{align}
    \mathbb{P}_0(t)=\mathcal{N}(t)\,\lhaf(\boldsymbol{\mathcal{A}}^{(0)}), \quad \mathbb{P}_1(t)=\mathcal{N}(t)\,\lhaf(\boldsymbol{\mathcal{A}}^{(1)}), \quad \mathbb{P}_2(t)=\frac{1}{2}\mathcal{N}(t)\,\lhaf(\boldsymbol{\mathcal{A}}^{(2)}),
\end{align}
where we defined the matrices
\begin{align}
    \boldsymbol{\mathcal{A}}^{(0)}=1,
    \quad
    \boldsymbol{\mathcal{A}}^{(1)}=
    \begin{pmatrix}
        \mathfrak{f}_1 & \mathfrak{a}_1 \\
        \mathfrak{a}_1 & \mathfrak{f}_2 \\      
    \end{pmatrix},
    \quad
    \boldsymbol{\mathcal{A}}^{(2)}=
    \begin{pmatrix}
        \mathfrak{f}_1 & \mathfrak{a}_1 & \mathfrak{a}_2 & \mathfrak{a}_1  \\
        \mathfrak{a}_1 & \mathfrak{f}_2 & \mathfrak{a}_1 & \mathfrak{a}_2 \\
        \mathfrak{a}_2 & \mathfrak{a}_1 & \mathfrak{f}_1 & \mathfrak{a}_1\\
        \mathfrak{a}_1 & \mathfrak{a}_2 & \mathfrak{a}_1 & \mathfrak{f}_2
    \end{pmatrix}, 
\end{align}
and where the normalization $\mathcal{N}$ is given by
\begin{align}
   \mathcal{N}(t)= \frac{\exp\left[-\frac{1}{2}
\bar{\boldsymbol{a}}_\br^\dagger (\boldsymbol{\Sigma}_\br-\frac{1}{2}\id_2)^{-1} \bar{\boldsymbol{a}}_\br\right]}{\sqrt{\det(\boldsymbol{\Sigma}_\br-\frac{1}{2}\id_2)}}.
\end{align}
Evaluating the loop-hafnians, we get the expressions cited in the main text, namely,
\begin{align}
    \mathbb{P}_0(t)=\mathcal{N}(t),\quad\mathbb{P}_1(t)=\mathbb{P}_0(t)\left[\mathfrak{a}_1+\mathfrak{f}_1\mathfrak{f}_2\right],\quad\mathbb{P}_2(t)=&\frac{1}{2}\mathbb{P}_0(t)\left[\left(\mathfrak{a}_2+\mathfrak{f}_2^2\right)\left(\mathfrak{a}_2+\mathfrak{f}_1^2\right)+4 \mathfrak{f}_1 \mathfrak{f}_2 \mathfrak{a}_1+2\,\mathfrak{a}_1^2\right].
\end{align}

\section{Explicit calculation of the zero time-delay second-order correlation function for general Gaussian states}\label{app:second_order_correlation_appendix}
Here we compute the $g^{(2)}(0)$ function for the reduced Gaussian state given by Eq.~\eqref{eq:gaussian_state_ladder_operator}. From Eq.~\eqref{eq:general_bar_second_order_zero_time_delay}, we know that when the detector state starts in the ground state, the second order coherence accrued in the detector matches exactly that of the incident radiation. Thus, it suffices to evaluate the expectation values $\langle a^\dagger a \rangle$ and $\langle a^{\dagger2}  a^2\rangle$ for the general radiation state 
\begin{align}
    \rho^{}_\gr =  \hat{D}_\alpha\,\hat{S}_\xi\,\rho^{}_\text{th}(\bar{n})\,\hat{S}^\dagger_\xi\,\hat{D}^\dagger_\alpha.
\end{align} 
The normal-ordered characteristic function corresponding to this state is given by 
\begin{align}
\label{eq:normal_ordered_characteristic function}
    \chi^{}_1(\zeta)=\exp\left\{\frac{1}{2} \!\left[2 |\alpha|  (\zeta -\zeta^*)-e^{-i \theta } (2 \bar{n}+1) \left(\zeta  \sinh r + e^{i \theta } \zeta^* \cosh r \right)\! \left(\zeta  \cosh r + e^{i \theta } \zeta^* \sinh r \right)+\zeta  \zeta^* \right]\right\},
\end{align}
where $\zeta\in\mathbb{C}$. Then, differentiating with respect to $\zeta$ and $\zeta^*$, through Eq.~\eqref{eq:s_ordered_moments} we arrive at
\begin{align}
\label{eq:total_number_of_gravitons}
    \langle\hat{a}^\dagger\hat{a}\rangle= |\alpha|^2+\left(\frac{1}{2}+\bar{n}\right) \cosh (2 r)-\frac{1}{2},
\end{align}
and
\begin{align}
    \langle\hat{a}^{\dagger 2}\hat{a}^2\rangle \!=\!\frac{1}{8} \!\left[8 |\alpha| ^2\! \left(|\alpha| ^2\!-2\right)\!-\!8 |\alpha| ^2 (2 \bar{n}+1) \!\cos (\theta ) \sinh (2 r)\!+\!8\! \left(2 |\alpha| ^2\!-\!1\!\right)\! (2 \bar{n}+1)\! \cosh(2r)\! + \! 3 (2 \bar{n}+1)^2 \!\cosh(4 r)\!+\!4 \bar{n} (\bar{n}+1)\!+\!5\right].
\end{align}
Combining the equations above, we find the result \eqref{eq:g2_general_gaussian_state} quoted in the main text.

\subsection{Effect of detector thermal noise in the determination of the zero time-delay second-order correlations}
To study how thermal noise affects the determination of the second-order coherence, we compute the $g_\br^{(2)}(0)$ for a detector initially in a thermal state with average thermal occupation number $n_\text{th}$. From the general expressions of the time evolution determined in App.~\ref{app:gaussian_evolution}, that is, from Eq.~\eqref{eq:evolution_general_covariance_matrix} and Eq.~\eqref{eq:evolution_general_displacement_vector}, we find that the state of the detector is specified by Eq.~\eqref{eq:evolved_covariance_and_displacement}, except that now
\begin{align}
\boldsymbol{\sigma}_\br(0)=
\begin{pmatrix}
    n_\text{th}+\frac{1}{2} & 0\\
    0 & n_\text{th}+\frac{1}{2} \\
\end{pmatrix}.
\end{align}
Then, we can determine $g_\br^{(2)}(0)$ exactly as above, yielding
\begin{align}
\label{eq:second_order_correlation_noisy_detector}
    g_{\br}^{(2)}(0)=\frac{\mathcal{I}_1(t)+\mathcal{I}_2(t)+\mathcal{I}_3(t)}{8 \left[n_\text{th}+\sin ^2(\gamma_\g t) \left(n_\gr - n_\text{th}\right)\right]^2},
\end{align}
where
\begin{align}
\begin{gathered}
    \mathcal{I}_1(t)=4 (2 \bar{n}+1) \cosh (2 r) \sin ^2(\gamma_\g t) \left[2 |\alpha| ^2+\left(2 n_\text{th}+1-2 |\alpha| ^2\right) \cos (2 \gamma_\g t)+2 n_\text{th}-1\right],\\
    \mathcal{I}_2(t)=\sin ^4(\gamma_\g t) \left[8 |\alpha| ^4-8 |\alpha| ^2 (2 \bar{n}+1) \cos (\theta ) \sinh (2 r)+3 (2 \bar{n}+1)^2 \cosh (4 r)+(2 \bar{n}+1)^2\right],\\
    \mathcal{I}_3(t)=4 \left[(2 n_\text{th}+1) \cos ^2(\gamma_\g t)-1\right] \left[(2 n_\text{th}+1) \cos ^2(\gamma_\g t)+4 |\alpha| ^2 \sin ^2(\gamma_\g t)-1\right].
\end{gathered}
\end{align}
In the limit $n_\text{th}\to0$ of Eq.~\eqref{eq:second_order_correlation_noisy_detector}, the time dependence drops and we recover the expression in Eq.~\eqref{eq:general_bar_second_order_zero_time_delay}. In the limit $\bar{n}\to0$, which corresponds to a displaced squeezed gravitational wave, we find
\begin{align}
    g_{\br}^{(2)}(0)=2-\frac{ \left[|\alpha|^2-\frac{1}{2}\sinh(2r)\right]^2\sin^4(\gamma_\g t)}{ \left[ n_\text{th} +\sin ^2(\gamma_\g t)(|\alpha| ^2+\sinh^2r- n_\text{th})\right]^2}.
\end{align}

The calculation above reveals how imperfections in the state preparation of the detector affect our ability to learn the statistical properties of the gravitational signal. In general, however, we also have to account for noise during the measurement itself. Using the evolution determined by the Markovian diffusive model defined by Eq.~\eqref{eq:open_dynamics}, which for a detector initialized to its ground state  corresponds to the covariance matrix and displacement vector defined in Eq.~\eqref{eq:open_evolution_bar}, we find that the second-order correlation function is
\begin{align}
\label{eq:g2_open_dynamics}
    g_{\br}^{(2)}(0)=2+\frac{ \sin ^4(\gamma_\g t) \left\{(2 \bar{n}+1) \sinh (2 r) \left[(2 \bar{n}+1) \sinh (2 r)-4 |\alpha|^2\right]-4 |\alpha|^4\right\}}{2\left[ \bar{N}(e^{\kappa  t}-1)+n_\gr \sin ^2(\gamma_\g  t)\right]^2}.
\end{align}
As expected, in the limit $t\to\infty$ of Eq.~\eqref{eq:g2_open_dynamics}, $g_\br^{(2)}\to2$. Assuming $\kappa=\omega/Q\ll1$, where $Q$ is the resonator's quality factor, we arrive at the condition $ k_\text{B}T \,t/\hbar Q=\Gamma_\text{th} \,t< n_\gr (\gamma_\g  t)^2$, where $k_\text{B}$ is Boltzmann's constant, $T$ the temperature of the thermal bath, and we used the approximation $\bar{N}\approx k_\text{B} T/\hbar Q$. This condition corresponds to the requirement that less than one environmental phonon enters the resonator throughout the duration of the experiment. 

\section{Explicit calculation of the first and second order terms in the homodyne correlation measurement scheme}
\label{app:field_correlations}
Here we compute explicitly the contributions appearing in Eqs.~\eqref{eq:first_order_delta} and \eqref{eq:second_order_delta}. The zeroth order term \eqref{eq:zeroth_order_delta} was already computed in the App.~\ref{app:second_order_correlation_appendix} and corresponds to the unnormalized second order coherence \eqref{eq:general_bar_second_order_zero_time_delay}, as is mentioned in the main text. Let us start from the first-order contribution. We have,
\begin{align}
    \langle:\!\Delta\hat{h}_{\phi}\Delta\hat{n}_\gr\!:\rangle=&\langle:\!\hat{h}_\phi\,\hat{a}^\dagger\hat{a}\!:\rangle-\langle \hat{h}_\phi \rangle\langle \hat{a}^\dagger\hat{a}\rangle\\
    =&\frac{1}{\sqrt{2}}\left(e^{-i\phi}\langle \hat{a}^\dagger \hat{a}\,\hat{a}\rangle+e^{i\phi}\langle \hat{a}^\dagger\hat{a}^\dagger\hat{a}\rangle\right)-\langle \hat{h}_\phi \rangle\langle \hat{a}^\dagger\hat{a}\rangle,
\end{align}
which, appealing again to the moment generating formula Eq.~\eqref{eq:s_ordered_moments} and the normal-ordered characteristic function \eqref{eq:normal_ordered_characteristic function}, yields
\begin{align}
\begin{gathered}
    \langle \hat{a}^\dagger \hat{a}\, \hat{a}\rangle = \frac{1}{2}  \left[2 \alpha ^* \left(| \alpha | ^2+(2 \bar{n}+1) \cosh (2 r)-1\right)-\alpha  e^{-i \theta } (2 \bar{n}+1) \sinh (2 r)\right],\\
    \langle \hat{a}^\dagger \hat{a}^\dagger \hat{a}\rangle= \frac{1}{2}  \left[2 \alpha  \left(| \alpha | ^2+(2 \bar{n}+1) \cosh (2 r)-1\right) - \alpha^* e^{i \theta } (2 \bar{n}+1)  \sinh (2 r)\right].
\end{gathered}
\end{align}
Therefore, we find
\begin{align}
\label{eq:quadrature_number_correlations}    \langle:\!\Delta\hat{h}_{\phi}\Delta\hat{n}_\gr\!:\rangle=\frac{1}{\sqrt{2}}\left(\alpha\, e^{-i\phi}+\alpha^* e^{i\phi}\right)\left[\left(\frac{1}{2}+\bar{n}\right)\cosh(2r)-\frac{1}{2}\right]-\frac{1}{\sqrt{2}}\left(\alpha\, e^{-i(\phi+\theta)}+\alpha^* e^{i(\phi+\theta)}\right)\sinh (2r)\left(\frac{1}{2}+\bar{n}\right).
\end{align}
From Eq.~\eqref{eq:quadrature_number_correlations}, one can see that this quantity vanishes for pure displaced states $|\alpha\rangle$, pure squeezed states $|\xi\rangle$, thermal states $\rho_\text{th}(\bar{n})$, mixed states with zero displacement, and pure Fock states. In all other cases, the quantity \eqref{eq:quadrature_number_correlations} is non-vanishing. Proceeding in a similar manner for the second-order term \eqref{eq:second_order_delta}, we find
\begin{align}
\langle:\!(\Delta\hat{h}_{\phi})^2\!:\rangle&=\langle:\!\hat{h}_\phi^2\!:\rangle-\langle\hat{h}_\phi\rangle^2\\
&=\frac{1}{2}\left(e^{-2i\phi}\langle  \hat{a}\,\hat{a}\rangle+e^{2i\phi}\langle \hat{a}^\dagger\hat{a}^\dagger\rangle +2\langle\hat{a}^\dagger\hat{a}\rangle\right)-\frac{1}{2}\left(\alpha \,e^{-i \phi} + \alpha^*e^{i \phi}\right)^2
\end{align}
which, finally, leads to the expression
\begin{align}
    \langle:\!(\Delta\hat{h}_{\phi})^2\!:\rangle=\left(\frac{1}{2}+ \bar{n}\right) \left[ \cosh (2 r)- \sinh (2 r) \cos (\theta -2 \phi )\right]-\frac{1}{2}.
\end{align}

\end{document}